\documentclass[conference]{IEEEtran}
\usepackage{graphicx}
\usepackage{xspace}

\usepackage{caption}
\usepackage{subcaption}
\usepackage{transparent}
\usepackage{mathtools}
\usepackage{comment}
\usepackage{balance}
\usepackage[hyphens]{url}
\usepackage{bbm}
\usepackage{array,longtable}
\usepackage{graphicx}
\usepackage{times}
\usepackage{crossreference}
\usepackage{tikz}
\usepackage{booktabs,makecell}
\usepackage{eurosym}
\usepackage{array}
\usepackage{multirow}
\usepackage{hyphenat}
\usepackage[section]{placeins}
\usepackage{listings}
\usepackage{footnote}

\usepackage{booktabs}
\usepackage{array,longtable}
\makesavenoteenv{tabular}
\makesavenoteenv{table}

\begin{document}

\title{A Longitudinal Analysis of Online Ad-Blocking Blacklists}

 \author{\IEEEauthorblockN{Saad Sajid Hashmi}
\IEEEauthorblockA{
Macquarie University\\
saad.hashmi@hdr.mq.edu.au}
\and
\IEEEauthorblockN{Muhammad Ikram}
\IEEEauthorblockA{Macquarie University\\
University of Michigan\\
muhammad.ikram@mq.edu.au}
\and
\IEEEauthorblockN{Mohamed Ali Kaafar}
\IEEEauthorblockA{Macquarie University\\
Data61, CSIRO\\
dali.kaafar@mq.edu.au}}

\maketitle

\begin{abstract}

Websites employ third-party ads and tracking services leveraging cookies and JavaScript code, to deliver ads and track users' behavior, causing privacy concerns. 
 To limit online tracking and block advertisements, several ad-blocking (black) lists have been curated consisting of URLs and domains of well-known ads and tracking services. 
  Using Internet Archive's Wayback Machine in this paper, we collect a retrospective view of the Web to analyze the evolution of ads and tracking services and evaluate the effectiveness of ad-blocking blacklists. %
  We propose metrics to capture the efficacy of ad-blocking blacklists to investigate whether these blacklists have been \textit{reactive} or \textit{proactive} in tackling the online ad and tracking services. We introduce a \textit{stability} metric to measure the temporal changes in ads and tracking domains blocked by ad-blocking blacklists 
  and a \textit{diversity} metric to measure the ratio of new ads and tracking domains detected. %
 We observe that ads and tracking domains in websites change over time, and among the ad-blocking blacklists that we investigated, our analysis reveals that some blacklists were more informed with the existence of ads and tracking domains,  %
 but their \textit{rate of change} was slower than other blacklists. %
  Our analysis also shows that Alexa top 5K websites in the US, Canada, and the UK have the most number of ads and tracking domains per website, and have the highest \textit{ proactive} scores. This suggests that ad-blocking blacklists are updated by prioritizing ads and tracking domains reported in the popular websites from these countries.
\end{abstract}

\section{Introduction}
\label{sec:intro}
For almost as long as the commercial world wide web has existed, third-party advertisements (in short ads) and web tracking practices have been used to support free services~\cite{Lerner2016}. %
To this end, first-party websites leverage various techniques such as third-party \texttt{iframes} and JavaScript to show ads and track users' activities on websites. %
While this helps to support the business model of most content providers, its use in showing ads and profiling users' activities raises serious privacy concerns. Besides, many third-party services such as ads are being used as \textit{malvertisements}~\cite{ikram2019chain, adsnetworksmalware}. %

A number of studies have characterized and measured the ads and tracking ecosystem to protect user privacy and limit intrusive ads, resulting in numerous ad-blocking tools such as Ghostery~\cite{ghostery}, Adblock~\cite{adblock}, Adblock Plus~\cite{adblockplus}, Disconnect~\cite{disconnect}, and Privacy Badger~\cite{privacybadger} for the web~\cite{Ikram2017} and mobile platforms~\cite{ikram2017first}. Most of these tools are fueled by community-driven public blacklists (such as EasyPrivacy~\cite{easyprivacy}, EasyList~\cite{easylist}, FanboyList~\cite{fanoby}, and hpHosts~\cite{hpHost}) and are evaluated for their (in)effectiveness~\cite{wills2016ad, Ikram2017}. Although the research work in this domain has been on-going for some time and ad-blocking tools have been developed, we believe these studies typically consist of short-term measurements of specific tracking techniques and the ad-blocking tools have not been comprehensively evaluated over time. Given that the online ads, web tracking, and prevention landscape is continuously evolving, studies performed at a snapshot or longitudinally starting in the present may not comprehensively illuminate online ads/tracking to improve ad-blocking. %

To fill the gap, we leverage the historical data extracted from the Internet Archive's Wayback Machine~\cite{wayback} to analyze online ads (resp. web tracking) and ad-blocking blacklists (listed in Table~\ref{tab:blacklist}). %
From online ads and web tracking perspective, we look at the prevalence and the evolution of ads and tracking domains since 2009. %
From an ad-blocking and tracking prevention perspective, we measure the longitudinal performance of various ad-blocking blacklists with respect to their {\it rate of change} and {\it update speed}. %
This paper makes the following main contributions:

    \noindent {\bf Prominence of a\&t domains.} 
     We study the evolution of the number of ad and tracking (a\&t for short) domains embedded in websites over time. Globally, since 2010 the presence of a\&t domains has been dynamic on an average of 55\% of the websites crawled, with 45\% exhibiting an annual increase and 10\% of the websites exhibiting an annual decrease in the number of a\&t domains embedded.   
     We observe a bias in the presence of a\&t domains across different countries. %
     with the Alexa top-5K websites in the US, Canada, and the UK exhibiting 
    the highest number of a\&t domains per website, whereas China and Iran had the least. %
     We also examine the number of a\&t services detected by each of the analyzed ad-blocking blacklist ({\it see} Table~\ref{tab:blacklist}). We observe that the blacklists such as Mahakala and hpHosts report the highest number of a\&t services while Cybercrime and Sa-blacklist detect the least. 
    
    \noindent {\bf Rate of change of tracking prevention lists.} %
    To measure the churn of third-party a\&t domains within a blacklist, we introduce two new metrics: stability and diversity. 
    Although each year, the ad-blocking blacklists mostly detect a\&t domains that have already been reported in the preceding years, they continue to detect new third-party a\&t services in the Alexa top 5K websites. We observe a gradual drop of new a\&t services detected by ad-blocking blacklists: on average from 19\% in 2011 to 10\% in 2017. %

\noindent    {\bf Update speed of ad-blocking blacklists.} We introduce two new metrics that measure the speed of updates of ad-blocking blacklists by (de)listing third-party a\&t domains. Most of the ad-blocking blacklists have {\it proactively} detected a\&t services since 2009. %
    Among the analyzed ad-blocking blacklists, we found that AdZHosts has the highest average {\it reactive score} of 0.6 and Mahakala has the highest average {\it proactive score} of 0.49.%

\section{Ads and Tracking on Web}
\label{sec:background}

We briefly discuss the mechanism of third-party online a\&t services and overview ad-blocking blacklists. %

\textbf{Online ads and web tracking:} %
Many online businesses and their revenue depend on online advertisements delivered by third-parties. With the advent of new technologies and platforms, third-party online a\&t services (resp. websites' publishers) have adopted new ways to deliver ads and track users' interaction. In Web platform, webpages isolate advertisements from content by placing ads in an {\it iframe}, where ad and tracking content hosted in the iframe is isolated from the hosting webpage, and browsers allow only specific cross frame interactions~\cite{barth2009securing}. The website publisher includes the ad and tracking components (i.e., JavaScript codes and cookies belonging to a third-party ad and tracking service) in their websites and allocate iframes to display ads. The website publisher is paid by the ad provider based on the number of ad clicks or impressions, or both. Most often, these techniques track users across websites for targeted ads---delivering ads based on their browsing/search or purchase history---revealing sensitive information about website users. %

Ads and tracking services operate by storing the user's personal identifiable information (PII) in the form of browser cookies, HTML5 local storage and flash cookies which is then shared with (other) third-parties either implicitly via HTTP referrer or explicitly through tracker-provided JavaScript programs. If the user is tracked within the website, only by the first party domain (website directly visited by the user), it is termed as {\it first-party tracking}. An application of first-party tracking is to count the repeated visits of a user to the website for {\it analytics}. If the user is tracked on the website by another domain, this is termed as {\it third-party tracking}. For example, if the user visits {\tt {bbc.com}}, and the advertisement on {\tt {bbc.com}} is fetched from {\tt {advertiser.com}}, then the user's browser sends a request to {\tt {advertiser.com}} without the user being aware of it. In our study, we focus on the JavaScript programs containing third-party domains which are embedded in the websites that are popular according to Alexa websites ranking. Therefore, in this paper, the terms `ads and trackers' and `third-party ads and trackers' are used interchangeably to mean third-party ads and tracking domains.   

\textbf{Ad-blocking blacklists:}
To counter the privacy threat posed by third-party a\&t services, several ad-blocking tools have been developed. They include pop-up blockers, privacy preserving web proxies (like Privoxy~\cite{privoxy}, Proxomitron~\cite{proxmitron}, and Pi-hole~\cite{pihole}) to filter and block a\&t traffic at network-layer and ad-blocking browser extensions (e.g., Adblock Plus, Ghostery, and Disconnect) that prevent web resources from rendering in the browser~\cite{Ikram2017, ikram2017first}. %
The most popular of these tools are the ad-blocking browser extensions which rely on filter-lists (also referred to as tracking prevention lists or ad-blocking lists), comprising of sets of URLs and domains of a\&t services. %
These blacklists are either crowd-sourced with feedback from web users or maintained by ad-blocking tools developers. %
Table~\ref{tab:blacklist} overviews several well-known a\&t blacklists used by ad-blocking tools. %

\section{Dataset and Methodology}
\label{sec:methodology}

In this section, we provide an overview of our dataset and our measurement methodology. %

\subsection{Dataset} 
\label{subsec:data}
We aim to study the prevalence of a\&t services in popular websites worldwide. We choose two sets of popular websites: %
{\it (i)} Alexa top 5K global websites (5K); and {\it (ii)} Alexa top 5K websites in the following fourteen countries: Australia (AU), Brazil (BR), Canada (CA), China (CN), Germany (DE), Israel (IL), India (IN), Iran (IR), Russia (RU), Saudi Arabia (SA), Singapore (SG), Ukraine (UA), United Kingdom (UK), and United States (US). %

To conduct our longitudinal study, we use web data from the Internet Archive's Wayback Machine. Since 1996, the Wayback Machine has archived full websites, including JavaScript codes, style sheets, and any multimedia resources that it can identify statically from the site's content. We refer to a single capture of a webpage (resp. a\&t domains in an ad-blocking blacklist) as a {\it snapshot}. Wayback Machine mirrors past snapshots of these websites on its own servers. We scrape the HTML DOM (Document Object Model, where HTML elements are defined as objects) of Alexa %
top 5K global and country-wise websites. The purpose of scraping the HTML DOM is to extract all JavaScript codes (along with their sources) from the websites, which is then used for our analysis. Wayback Machine has a number of archived snapshots, with varying intervals, for each website. We utilize Memento API~\cite{momento} to capture snapshots at intervals of three months to detect most---if not all---of the ads and tracking domains that appeared on a website for a given year, during the period of 2009 to 2017. Memento API provides the nearest time-stamp for the archived snapshot of a website (resp. an ad-blocking blacklist) from the date provided. 

We empirically observed that 85\% of the snapshots are within an interval of 6 months. %
From these snapshots, we then obtain third-party sources from the embedded JavaScript codes, and by comparing it with the second level domains of ad-blocking blacklists, we obtain the a\&t domains in the crawled websites. %
For example, if {\tt {http://ad.doubleclick.net/dot.gif}} appears in the script tag of {\tt {sportsbet.com}}, then the second level domain ({\tt {doubleclick.net}}) is extracted from the URL, and compared against the domains in the ad-blocking blacklists. %

Similarly, by using Internet Archive's Wayback Machine, we obtain the snapshots of ad-blocking blacklists. Table~\ref{tab:blacklist} lists the number of distinct domains and the snapshots period of the analyzed ad-blocking blacklists. %

\begin{table}[!ht]
\caption{\small Overview of the analyzed ad-blocking blacklists.} %
\scriptsize
\centering
\tabcolsep=0.12cm
\begin{tabular}{l l l r}
\toprule
 & & {\bf Snapshots} & {\bf \# Distinct }\\  
{\bf Blacklist} & {\bf Description} &{\bf Period} & {\bf Domains}\\  \midrule
AdZHosts~\cite{adzhosts} &  Ad/tracker/malicious hosts & 2017-17 & 81,294\\

AdGuard~\cite{adguard} & Ad/tracker/phishing hosts & 2017-17 & 12,881\\

Cameleon~\cite{cameleon} & Ad/tracker hosts & 2009-16 & 47,390\\

CyberCrime~\cite{cybercrime} & Tracker/malware/phishing hosts & 2013-16 & 6,557\\

EasyList~\cite{easylist} & Ad/trackers/analytics & 2010-17 & 13,882\\

FanboySocial~\cite{fanoby} & Social buttons/widgets hosts & 2013-17& 282\\%

EasyPrivacy~\cite{easyprivacy} & Trackers/analytics hosts & 2011-17 & 7,050\\%

QuidsUp~\cite{QuidsUp} & Tracker/malicious hosts& 2017-17 & 865\\

EasyList\_China~\cite{EasyListChina} & Chinese EasyList websites & 2014-17& 18,639\\

hpHosts~\cite{hpHost} & Ad/tracker/malicious hosts & 2009-17 & 856,498\\

MVPs~\cite{mvpshosts}& Ad/tracker/malicious hosts & 2011-17 & 414,461\\
AdAway~\cite{adaway}& Mobile ad/tracker hosts & 2013-16 & 137\\%
Disconnect~\cite{disconnect} & Ad/tracker hosts & 2015-16 & 15,238\\
Mahakala~\cite{mahakala} & Ad/tracker hosts & 2015-17 & 1,846,198\\
Sa-blacklist~\cite{sa-filterlist} & Ad/tracker hosts & 2009-13 & 1,253,243\\
\bottomrule
\end{tabular}
\label{tab:blacklist}
\end{table}

\subsection{Methodology}
\label{subsec:methodology}
The objective of this work is three fold. We study the:
\textit{(i)} prominence of a\&t domains on the Alexa top 5K worldwide as well as top 5K websites in different countries, %
\textit{(ii)} rate of change of ad-blocking blacklists, and
\textit{(iii)} update speed of ad-blocking blacklists. 

\textbf{Prominence of a\&t domains:} For our first objective, we compare the second-level domains of JavaScript sources (observed in retrospectively crawled top 5K websites from the years 2009 to 2017) with those of known a\&t domains from the ad-blocking blacklists. This comparison reveals the number of a\&t domains blocked in different ad-blocking blacklists' yearly snapshots and the number of websites that embed them. %

By measuring the number of a\&t domains on the Alexa top 5K websites of different countries, we also measure the prevalence of a\&t domains in a given country's top websites. Besides detecting existing a\&t domains in the wild, an effective ad-blocking blacklist also has to evolve and detect new a\&t domains over time to counter the ever-changing tracking landscape. Therefore, we also investigate the number of new a\&t domains detected by the ad-blocking blacklists from 2009 to 2017.    

\textbf{Rate of change of ad-blocking blacklists:} 
For our second objective, we compare the a\&t domains detected in two consecutive years. This comparison reveals whether there has been any change in the a\&t domains blocked compared with the preceding year, and indicates the evolution of a\&t domains in the ad-blocking blacklist. %
To this end, we propose the \textit{stability} metric: %

 \begin{equation}
    Stability = \frac{\mid T_{i} \cap T_{i-1}\mid}{\mid T_{i} \cup T_{i-1}\mid},
    \label{eq:stability}
\end{equation}

where {${\bf T}_{i}$} and {${\bf T}_{i-1}$} are the sets of distinct a\&t domains detected in the current year, $i$, and the preceding year, $i-1$. %
Stability score of 1 implies that the list of a\&t domains blocked in a given year is identical to the list from the preceding year, and stability score of 0 implies that the list of a\&t domains blocked in a given year is completely different than the list from the preceding year. %
By default, the stability of any ad-blocking blacklist for the first year is always zero. 

From Eq.~\ref{eq:stability}, the stability score of an ad-blocking blacklist may be {\it low}, however, it can detect new a\&t domains, because the stability score considers the given year and the preceding year only. Therefore, if a\&t domains that were blocked in years before the preceding year are detected again, they have no impact on the measurement of stability. To overcome this limitation, we introduce the \textit{diversity} metric, which measures the ratio of {\it new} a\&t domains detected by the ad-blocking blacklist for a given year. The diversity of an ad-blocking blacklist for a given year is measured as: 
\begin{equation}
    Diversity = 1 - \frac{\mid T_{i} \cap T_a\mid}{\mid T_{i}\mid}\label{eq:diversity},
\end{equation}
where {${\bf T}_a$} is the set of distinct a\&t domains detected in all previous years.
So a diversity of $1$ implies that all the a\&t domains blocked in a given year are new, i.e., they have not been detected by that ad-blocking blacklist in the previous years, and a diversity of 0 implies that none of the a\&t domains blocked in a given year are new. %
By default, the diversity score for any ad-blocking blacklist is always $1$ for the first year.

\textbf{Update speed of ad-blocking blacklists:} For our third objective, we study the speed by which each blacklist updated itself to encounter new threats. The \textit{time difference} is taken to be the number of days between the first occurrence of an a\&t domain in a website in the given year, and its addition to the blacklist. The time difference is taken to be positive if an a\&t domain is seen on the web before it gets added to the blacklist. It is negative otherwise. 
We consider the blacklist is reactive if the time difference is positive, and proactive otherwise. We aim to introduce the metrics for which a high score implies high performance. The performance of a blacklist is considered good if it detects more a\&t domains in the least amount of time or it contains a\&t domains prior to them being observed on the web. Therefore, to measure the reactive score, we divide the number of a\&t domains per website detected {\it reactively} by the average time difference, and compute proactive score by taking the product of the number of a\&t domains per website detected proactively and average time difference. The scores are then normalized between $0$ and $1$. The reactive and proactive scores of an ad-blocking blacklist for a given year are measured as:
\begin{equation}
    Reactive = \;\frac{(\frac{\sum_{}^{domains} \mid T_{react}\mid}{t_{days}})}{Reactive_{max}}
    \label{eq:react}
\end{equation}
\begin{equation}
    Proactive =\;\frac{ (\sum_{}^{domains}\mid T_{proact}\mid\;) \;* \;\mid t_{days}\mid}{Proactive_{max}} \label{eq:proact}
\end{equation}

where \textit{$T_{react}$} and \textit{$T_{proact}$} are the set of distinct a\&t domains per website detected reactively and proactively, respectively, in the given year and \textit{$t_{days}$} is the average time difference in days to add those a\&t domains in blacklists. Since for proactive a\&t domains blocked the average time difference is negative, we only consider the magnitude and ignore the negative sign.

A low reactive (resp. proactive) score implies that the ad-blocking blacklist is either blocking low number of a\&t domains or it is taking a large amount of time to add a\&t domains in the blacklist (or in case of proactive score, adds a\&t domains just recently before being observed in the top 5K websites) or both. A high reactive (resp. proactive) score implies that the ad-blocking blacklist is either blocking large quantities of a\&t domains or it is taking less amount of time to add those a\&t domains in the list (or in case of proactive score, adds a\&t domains prior to when they are observed in the top 5K websites) or both. 

Note that if an ad-blocking blacklist started its service (or we capture its first snapshot from Wayback Machine) after the first occurrence of a given a\&t domain on the web, the reactive score is computed by taking the time difference from when the a\&t domains are added to a given ad-blocking blacklist. %
If an a\&t domain appears in the first version (or snapshot) of a given blacklist then %
we consider the time difference as $1$ instead of $0$ (to avoid division by $0$).   

Similarly, for a given ad-blocking blacklist, we compute the reactive and proactive scores per country per year. %
We then average the scores over the countries and years. The normalized reactive and proactive scores are then computed by dividing the score by the maximum observed score at the lowest aggregation level i.e., total a\&t domains per year per country. 

In Section~\ref{subsec:evolution}, we use the above four metrics for different countries. All these metrics take the set of a\&t domains as input. For the ad-blocking blacklists, that set of a\&t domains comprises the distinct a\&t domains blocked by the given ad-blocking blacklist. For countries, the set of a\&t domains consists of the distinct a\&t domains that occurred in ad-blocking blacklists on the Alexa top 5K websites of the given country. For example, the stability of a country measures the similarity in the a\&t domains detected (by the union of all the ad-blocking blacklists) on the Alexa top 5K websites of that country compared with the preceding year. Similarly, we use the other three metrics for the countries.

\section{Analysis and Results}
\label{sec:analysis}

{\it Next}, we present results of our analysis. We measure the prominence of a\&t domains to establish whether some countries' top 5K websites consistently had high number of a\&t domains over the period of time, and study whether websites added new a\&t domains each year to avoid ad-blocking tools. In order to understand the variation in effectiveness of ad-blocking blacklists in recent years, we investigate how the ad-blocking blacklists have performed in recent years in terms of number of a\&t domains detected on the Alexa top 5K websites, and their rate of change measured by {\it stability} and {\t diversity} metrics. We also investigate the update speed of ad-blocking blacklists in terms of reactive and proactive scores.

\subsection{Prominence of Ads and Tracking Domains}
\label{subsec:trackerbehaviour}
\begin{figure*}[bt!]
\centering
\begin{minipage}{0.24\textwidth}
\includegraphics[scale=0.30, keepaspectratio]{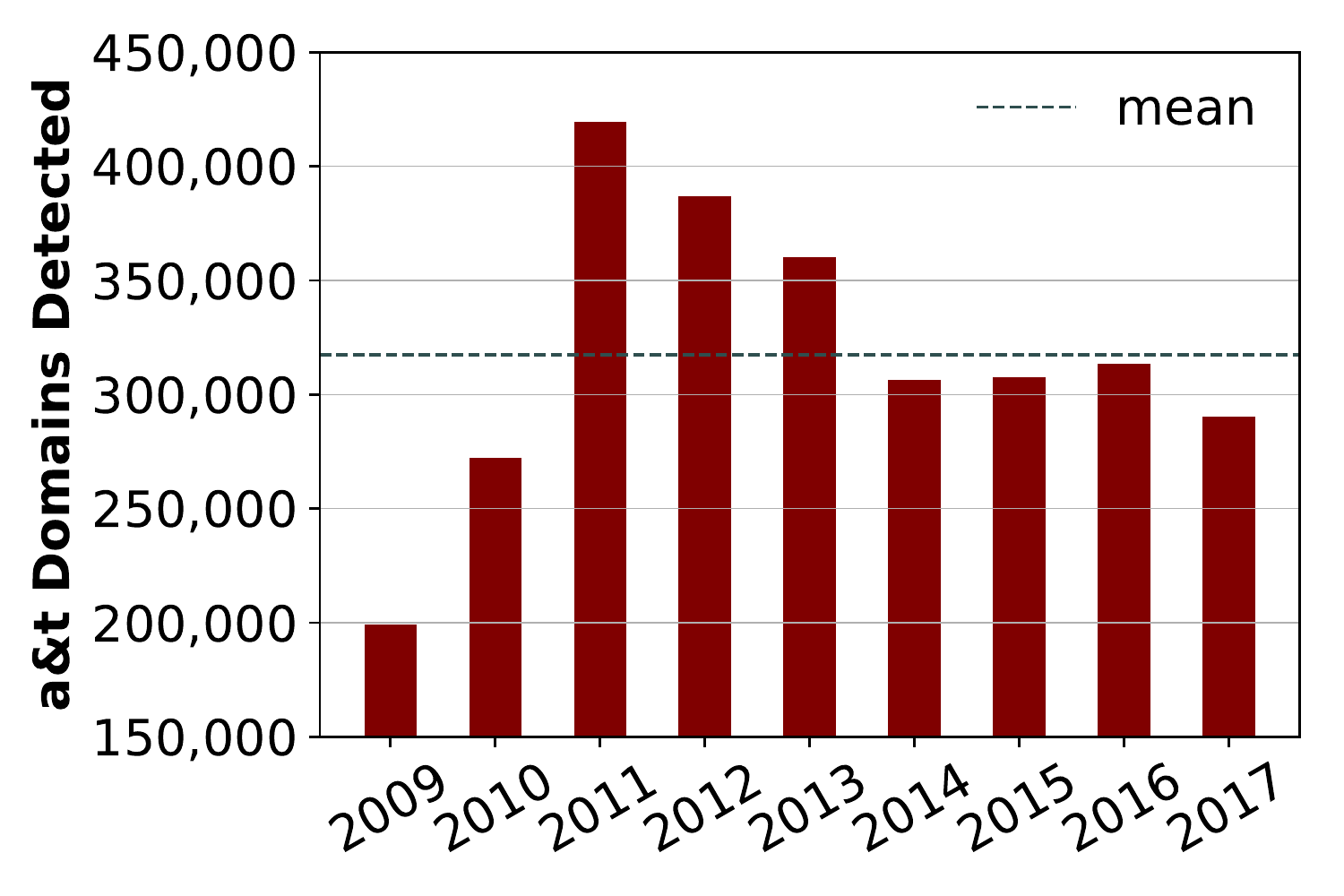}
\subcaption{}
    \label{fig:1(a)}
\end{minipage}
\begin{minipage}{0.24\textwidth}
\includegraphics[scale=0.30, keepaspectratio]{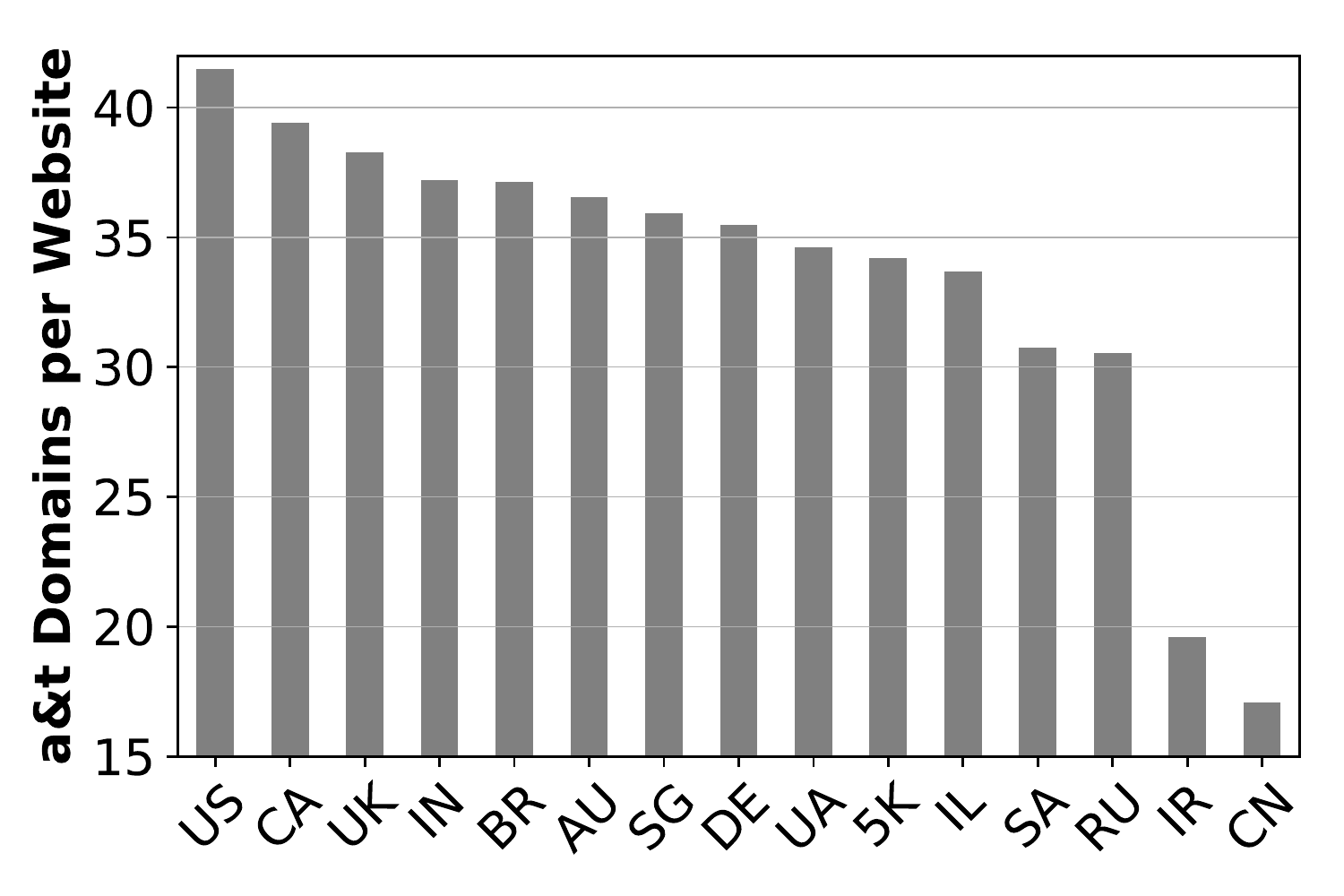}
\subcaption{}
    \label{fig:1(b)}
\end{minipage}
\begin{minipage}{0.24\textwidth}
\includegraphics[scale=0.30, keepaspectratio]{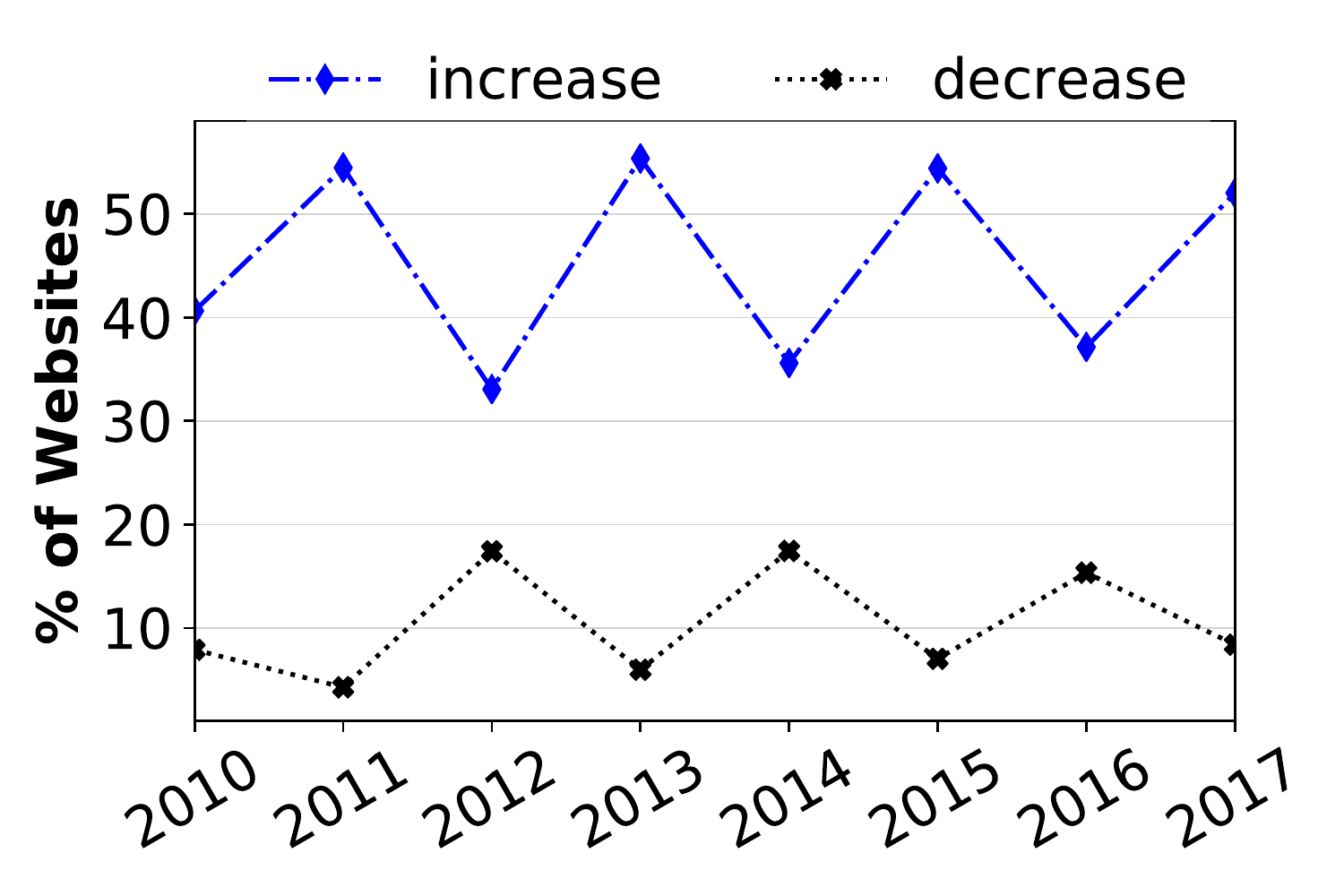}
\subcaption{}
    \label{fig:1(c)}
\end{minipage}
\begin{minipage}{0.24\textwidth}
\includegraphics[scale=0.30, keepaspectratio]{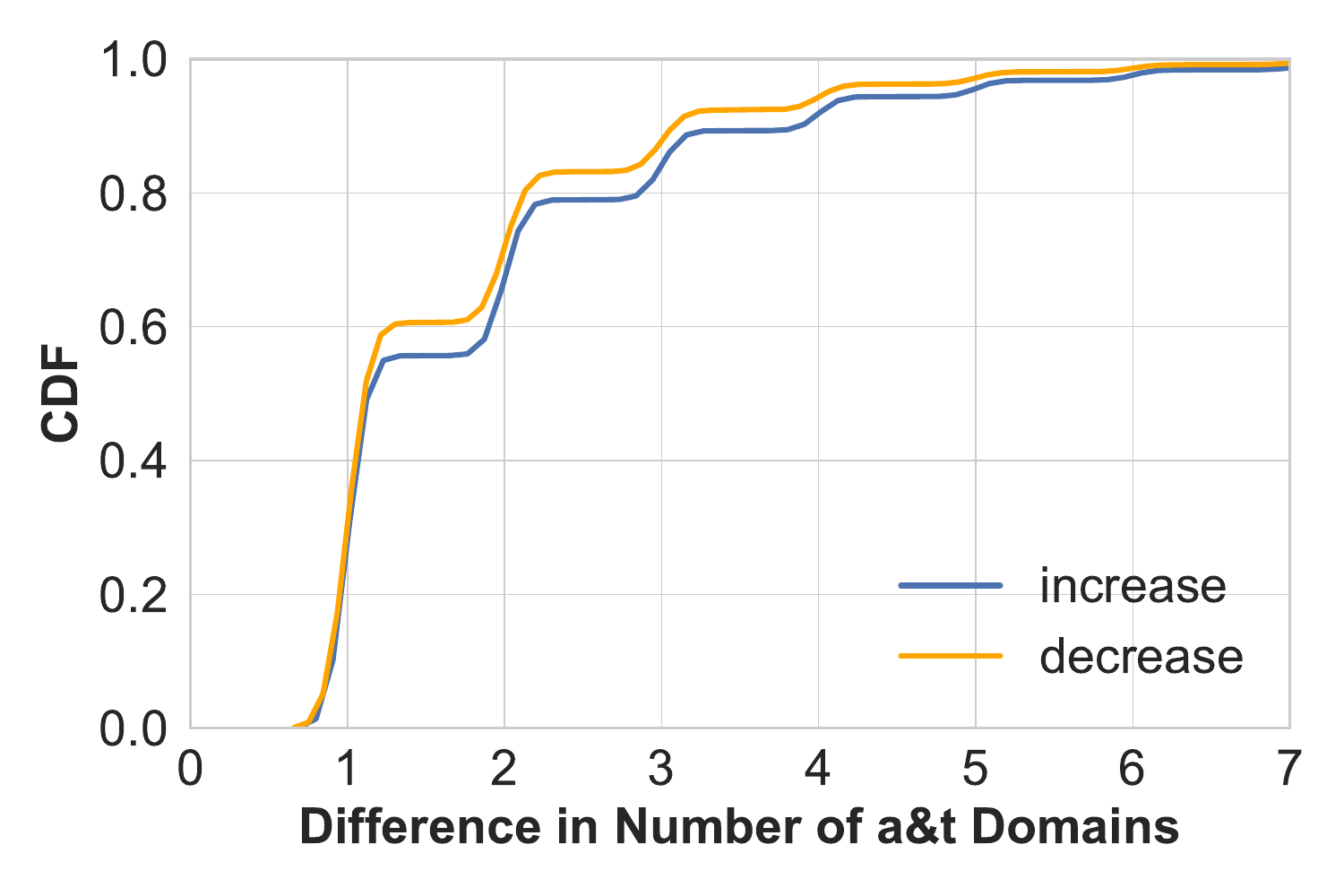}
\subcaption{}
    \label{fig:1(d)}
\end{minipage}
  \caption{\small (a) Annual sum of distinct a\&t domains per website (b) Distinct a\&t domains per website on the Alexa top 5K websites of each country (c) Annual Percentage of websites with an increase or decrease in number of a\&t domains from preceding year (d) Distribution of change in number of a\&t domains from the websites in Figure \ref{fig:1(c)}.}
  \label{fig:3}
\end{figure*}

 \textbf{Annual Measurement of a\&t domains:} From the crawled data, we measure the number of a\&t services observed in each year. Figure \ref{fig:1(a)} shows the number of a\&t domains blocked by ad-blocking blacklists. We observe that a\&t domains peak in 2011, before gradually declining to the mean number 317,357 of a\&t domains for all the years. One possible explanation for the decline after 2011 is that some websites embed new a\&t domains over time either because the a\&t domain is not functioning anymore or to evade ad-blocking tools~\cite{zhu2018measuring}. %
Also, all the new a\&t domains may not have been detected yet by the analyzed ad-blocking blacklists, and thus not reflected in the annual a\&t domains measurement. For example, we note that in 2011, {\tt {wikia.com}} embed {\tt {google-analytics.com}}, {\tt {quantserve.com}}, and {\tt {nocookie.net}} in its HTML whereas in 2016 {\tt {wikia.com}} had only {\tt {optimizely.com}} as a third-party a\&t domain.

\textbf{Country-wise measurement:} {\it Next}, we map the a\&t domains observed on the Alexa top 5K websites of different countries. We acknowledge that there are some limitations in capturing snapshots with Wayback Machine, and many websites from top 5K had to be discarded due to the challenges mentioned in \S \ref{sec:conc}. Therefore, the number of websites do not remain equal for all the countries, and instead of measuring the total number of a\&t domains in all the websites, we measure the average number of a\&t domains per website for each country. 
 
Figure~\ref{fig:1(b)} shows the union of distinct a\&t domains per website from 2009 to 2017. We observe that the US, Canada, and the UK have the most number of a\&t domains per website. %
As the ad-blocking blacklists are mostly updated through crowd-sourcing~\cite{vastel2018filters}, the prevalence of a\&t domains across countries also suggests that the %
crowd-sourcing users for the analyzed ad-blocking blacklists are more active in the countries reporting the most number of a\&t domains, compared to countries like Iran and China, that have detected the lowest number of a\&t domains per website. 

Figure \ref{fig:4(a)} shows the annual number of a\&t domains blocked on the Alexa top 5K websites of the top seven countries in Figure \ref{fig:1(b)}. Note that for ease of presentation, we select countries with most number of a\&t domains detected on the Alexa top 5K websites. The a\&t domains country-wise trend is similar to the overall trend in Figure \ref{fig:1(a)}, suggesting that the new a\&t domains may not yet have been detected by the analyzed ad-blocking blacklists. %

\textbf{Domain-wise measurement of a\&t domains:} Figure \ref{fig:1(c)} shows the percentage of websites with a change in the number of a\&t domains from the preceding year. We observe that, from 2010 to 2017, the presence of a\&t domains has been dynamic on an average of  55\%  of  the  analyzed websites,  with  45\%  exhibiting an  annual  increase  and  10\%  of  the  websites  exhibiting an  annual  decrease. %
Moreover, nearly 80\% of those changes are a difference of one or two a\&t domains (cf. Figure \ref{fig:1(d)}). 
For example, in {\tt {tomshardware.com}} there are four a\&t domains in 2009, ten in 2010, six in 2011, one in 2012, five in 2013, two in 2014, none in 2015, two in 2016, and one in 2017.

\textbf{Measurement of a\&t domains in blacklists:} Table~\ref{tab:blacklist_summary} lists the number of a\&t domains detected by the ad-blocking blacklists, and the mean time (in days) taken to add them in blacklist. The table also lists the number of a\&t domains detected reactively and proactively for each ad-blocking blacklist. The number of a\&t domains detected is the annual sum of distinct a\&t domains per website for the years 2009 to 2017. %
 We observe (cf. Table~\ref{tab:blacklist_summary}) that the number 2,460K of a\&t domains blocked by Mahakala is the highest. Not surprisingly, Mahakala also contains the most number 1,846K of distinct a\&t domains %
 We also observe that EasyList and EasyPrivacy contain 13,882 and 7,050 a\&t domains, respectively, and these two blacklists are able to detect 393K instances of those a\&t domains within the analyzed top 5K websites. Surprisingly, Sa-blacklist that contain 1,253K a\&t domains is only able to detect 65,698 instances of those a\&t domains. This reveals that EasyList and EasyPrivacy though {\it fairly lean} contain mostly {\it useful rules} to block a\&t domains in the top 5K websites. Conversely Sa-blacklist contains mostly {\it stale or otherwise useless rules} to block a\&t domains on the Alexa top 5K websites. CyberCrime blocks the least number 3,349 of a\&t domains, and one reason why this number is low is that a significant fraction of domains in CyberCrime belongs to malvertisements~\cite{CC_stats}. %

\begin{figure}[ht!]
\centering
\begin{minipage}{\columnwidth}
\includegraphics[scale=0.29, keepaspectratio]{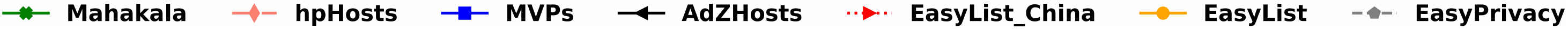}
\end{minipage}

\begin{minipage}{0.31\columnwidth}
\includegraphics[scale=0.195, keepaspectratio]{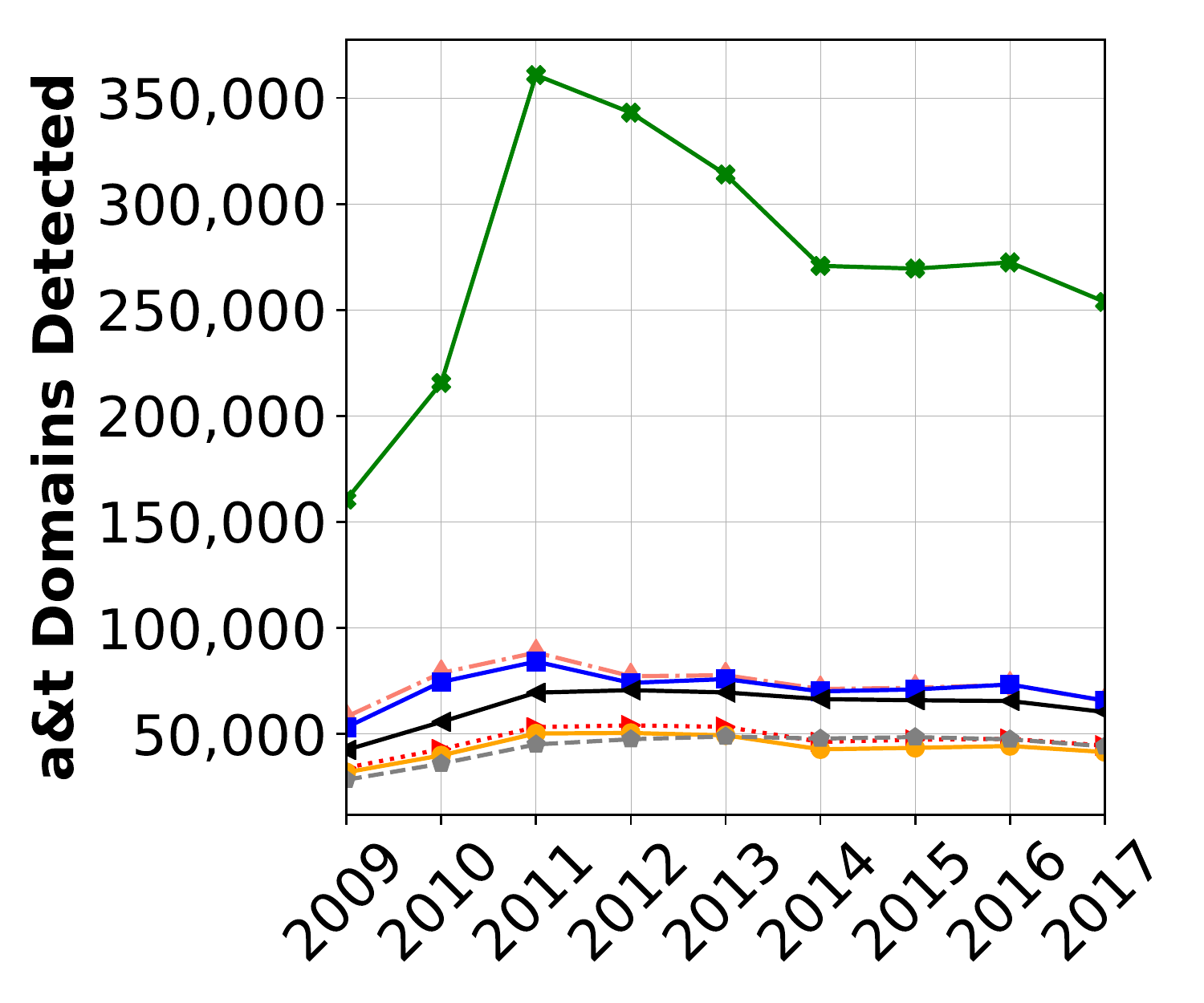}
\subcaption{}
    \label{fig:3(a)}
\end{minipage}
\begin{minipage}{0.31\columnwidth}
\includegraphics[scale=0.195, keepaspectratio]{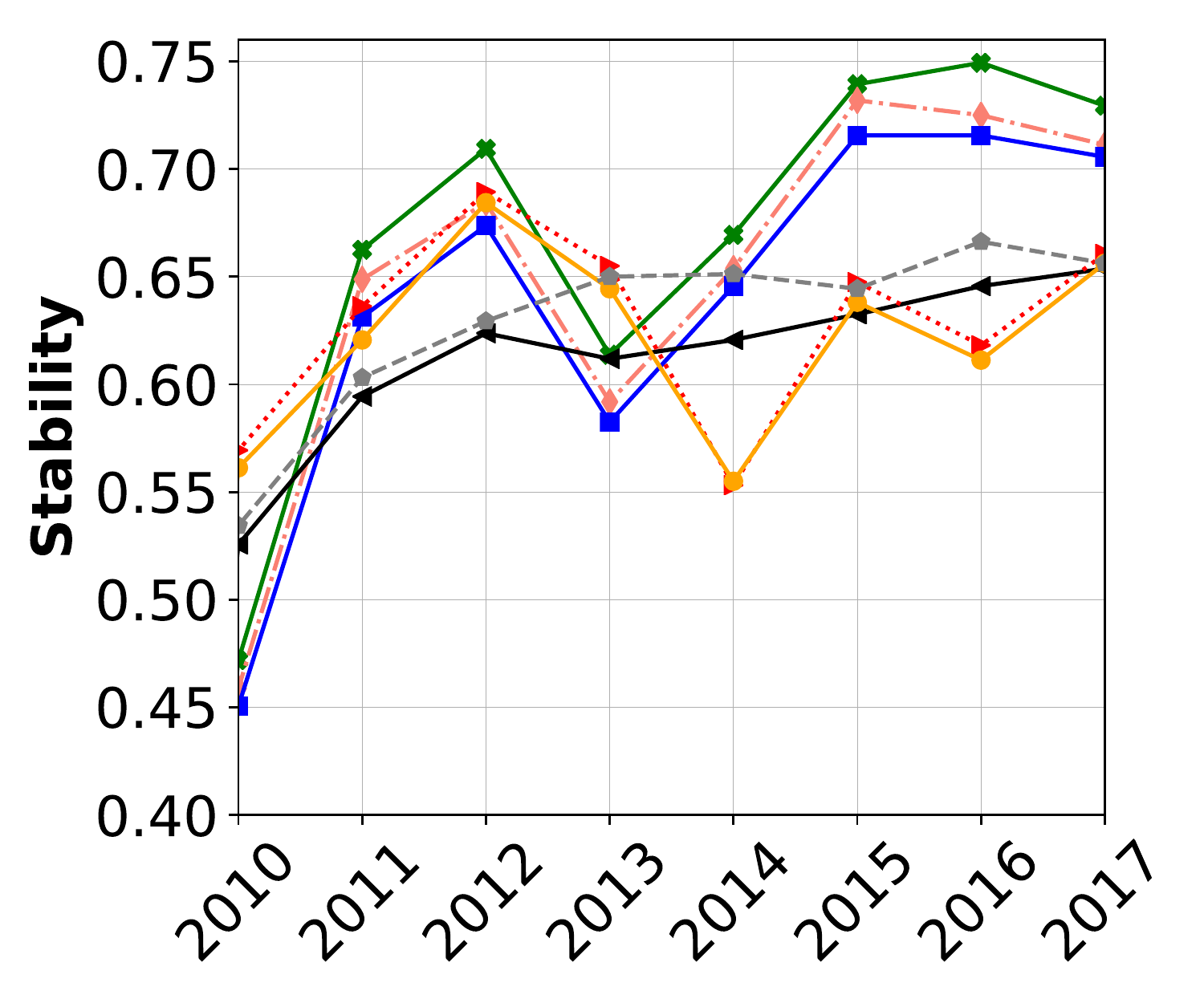}
\subcaption{}
    \label{fig:3(b)}
\end{minipage}
\begin{minipage}{0.32\columnwidth}
\includegraphics[scale=0.195, keepaspectratio]{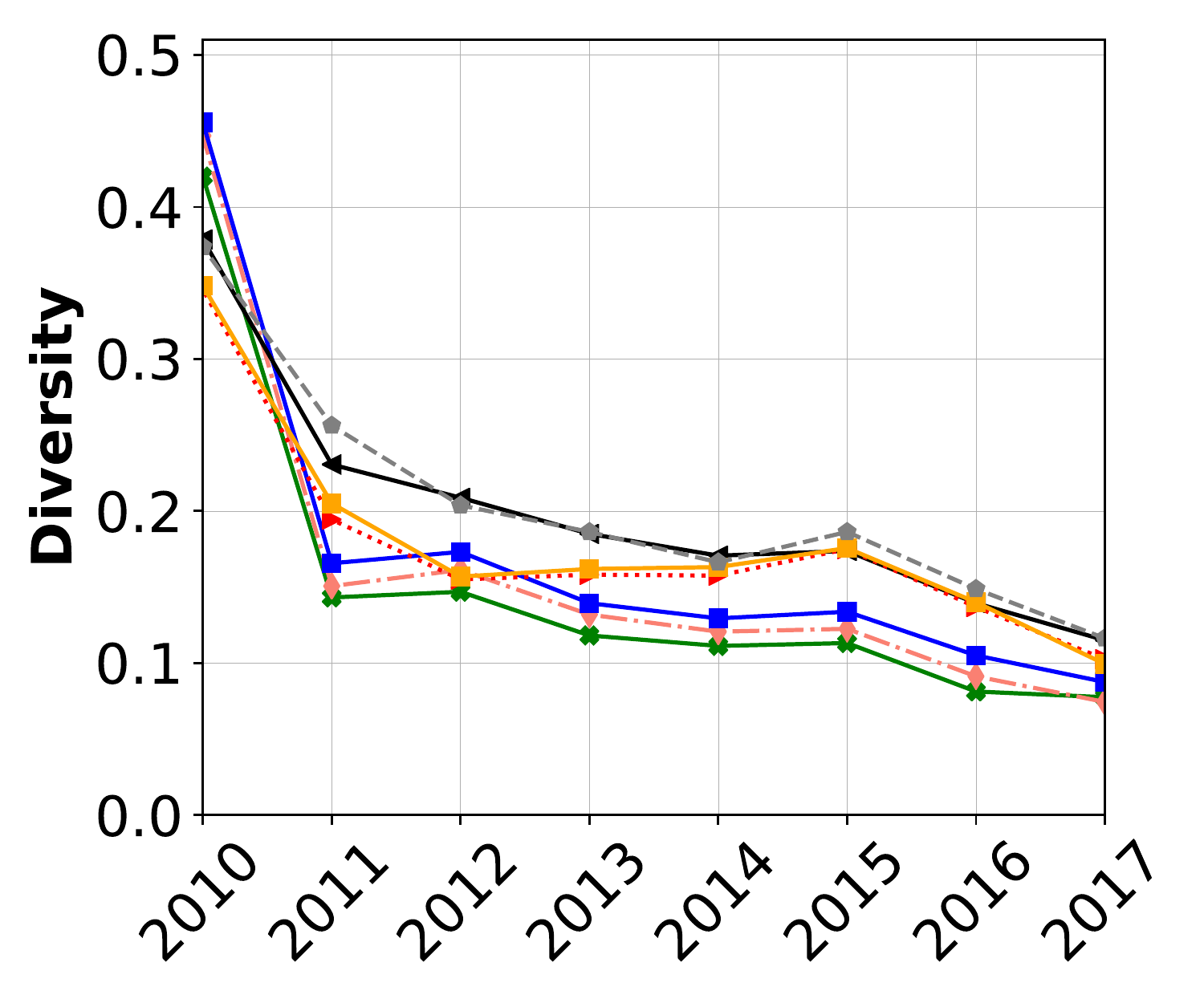}
\subcaption{}
    \label{fig:3(c)}
\end{minipage}
  \caption{\small (a) Annual a\&t domains detected by blacklists (b) Stability and (c)  Diversity of blacklists. With the gradual rise in stability and gradual decline in diversity, blacklists are blocking more a\&t domains similar to the preceding years, and less ratio of new a\&t domains.}
  \label{fig:3}
\end{figure}

Table~\ref{tab:blacklist_summary} lists 
ad-blocking blacklists ranked by the number of a\&t domains blocked on the Alexa top 5K websites. 
Annual number of a\&t domains blocked by the top seven ad-blocking blacklists in figure~\ref{fig:3(a)} indicates that the prevalence of a\&t domains peaks in 2011 for all the lists and then gradually declines and remains almost unaltered over the remaining period. This trend is similar to the number of a\&t domains observed in Figure \ref{fig:1(a)}. %
\begin{figure}[ht!]
    \centering
    \begin{minipage}{0.42\columnwidth}
        \includegraphics[scale=0.26, keepaspectratio]{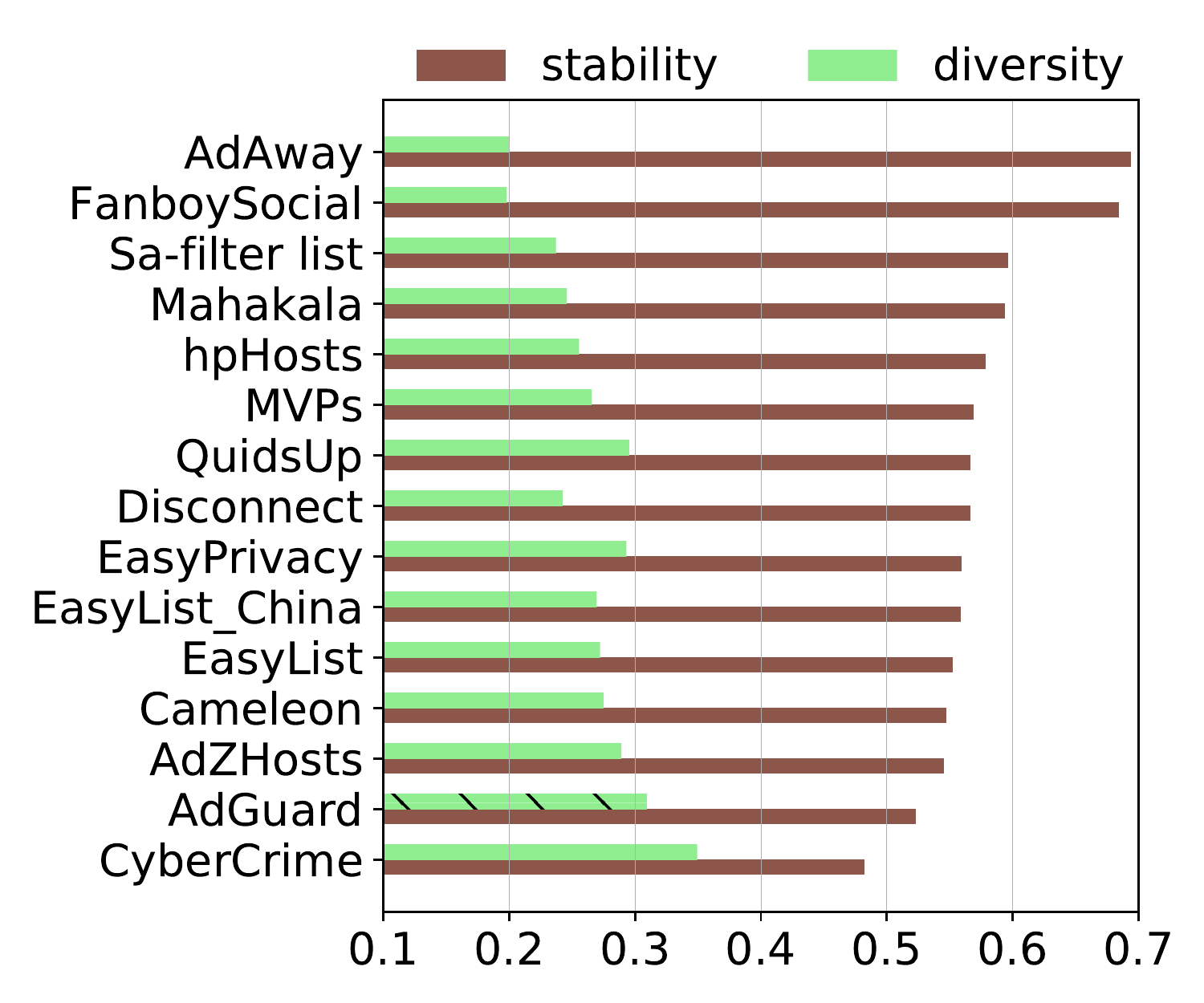}
        \subcaption{}
            \label{fig:2(a)}
    \end{minipage}
    \begin{minipage}{0.56\columnwidth}
        \includegraphics[scale=0.339, keepaspectratio]{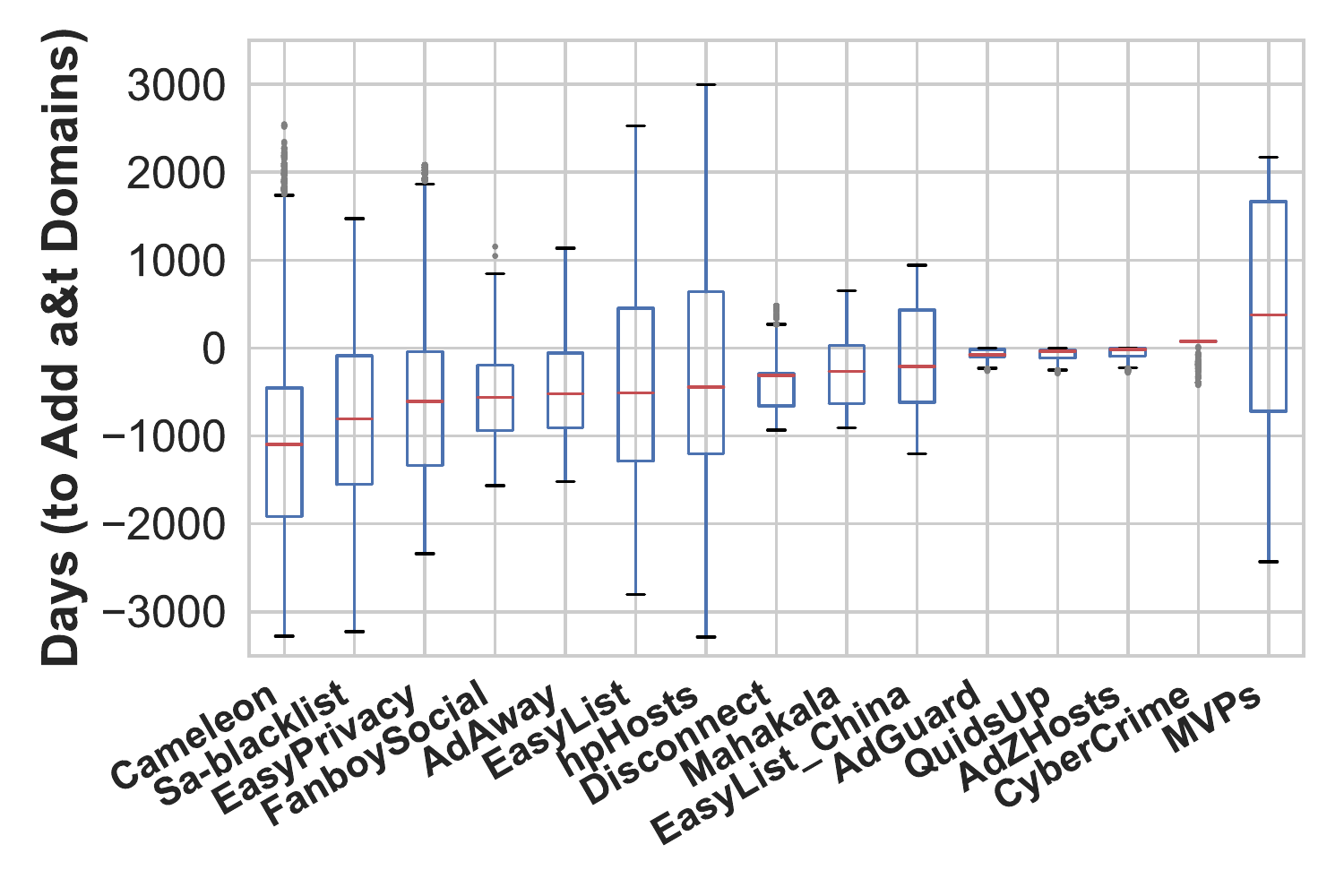}
        \subcaption{}
            \label{fig:2(b)}
    \end{minipage}
    \caption{\small (a) Average stability and diversity score of ad-blocking blacklists (b) Distribution of time taken (in days) to add a\&t domains in each blacklist. 
    The blacklists on left side are proactive whereas those on right side are reactive.}
    \label{fig:2}
\end{figure}
 
\begin{table}[!ht]
\caption{\small Number of a\&t domains detected and time taken (in days) to add them in ad-blocking blacklists. The number of a\&t domains detected is the annual sum of distinct a\&t domains per website for the years 2009 to 2017. Negative values for days indicate that the a\&t domains were added in blacklists before being observed on the Alexa top 5K websites.}
\scriptsize
\centering
\tabcolsep=0.1cm
\begin{tabular}{l l l l l l l}
\toprule
{\bf Black- } & {\bf Total a\&t} & {\bf Mean} & {\bf \# Reactive} & {\bf Mean} & {\bf \# Proactive} & {\bf Mean} \\  
{\bf List} & {\bf Domains} & {\bf Days} & {\bf a\&t Domains} & {\bf Days} & {\bf a\&t Domains} & {\bf Days} \\  
\midrule
Mahakala & 2,460,303 & -70 & 1,915,941 & 35 & 544,362 & -438 \\
hpHosts & 662,287 & -289 & 240,112 & 1,220 & 422,175 & -1,148 \\
MVPs & 642,023 & 294 & 399,207 & 1,093 & 242,816 & -1,020 \\
AdZHosts & 566,228 & -5 & 509,887 & 1 & 56,341 & -46 \\
EasyList\_China & 423,514 & -78 & 290,945 & 135 & 132,569 & -545 \\
EasyList & 393,990 & -381 & 150,384 & 815& 243,606 & -1,120 \\
EasyPrivacy & 393,914 & -510 & 143,316 & 321 & 250,598 & -985 \\
AdGuard & 386,220 & -3 & 370,247 & 1 & 15,973 & -80 \\
Cameleon & 365,851 & -1,159 & 28,337 & 489 & 337,514 & -1,297 \\
Disconnect & 199,773 & -93 & 155,389 & 7 & 44,384 & -441 \\
FanboySocial & 135,163 & -283 & 76,398 & 53 & 58,765 & -719 \\
QuidsUp & 116,297 & -8 & 101,852 & 1 & 14,445 & -68 \\
AdAway & 79,861 & -290 & 47,370 & 25 & 32,491 & -749 \\
Sa-blacklist & 65,598 & -810 & 14,117 & 546 & 51,481 & -1,182\\
CyberCrime & 3,349 & 10 & 3,214 & 22 & 135 & -257 \\

\bottomrule
\end{tabular}
\label{tab:blacklist_summary}
\end{table}

\subsection{Rate of Change}
\label{subsec:evolution}
\textbf{Stability and diversity analysis of blacklists:} Figure \ref{fig:2(a)} shows the average {\it stability} and {\it diversity} scores of the ad-blocking blacklists in detecting a\&t domains in the top 5K global as well as the top 5K websites in different countries from 2009 to 2017. We observe that the ad-blocking blacklists having high stability scores tend to have low diversity scores and vice-versa. For instance, AdAway and Fanboy Social have highest average stability and lowest average diversity, whereas CyberCrime and AdGuard have lowest average stability and highest average diversity. The average stability scores of top seven ad-blocking blacklists (in Table \ref{tab:blacklist_summary}) range between 0.5 and 0.6, and the average diversity scores are in the range of 0.2 to 0.3. %

The annual average stability and diversity scores of the ad-blocking blacklists in Figure~\ref{fig:3(b)} and \ref{fig:3(c)} show the gradual rise in stability and the gradual decline in diversity of ad-blocking blacklists, respectively. The stability is $0$ and diversity is $1$ in 2009. 
In 2010, we observe that the set of distinct a\&t domains blocked by the ad-blocking blacklists is significantly different from the one in 2009. This is evident from the stability and diversity scores, which are at their lowest and highest peaks, respectively. Since the diversity score is less than 0.5, it implies that the set of distinct a\&t domains blocked in 2010 had less than half the a\&t domains that were newly detected in 2010. 
We also observe a sharp increase in stability and steep decline in diversity in 2011, and no significant variation in the later years. This suggests that although the annual set of a\&t domains blocked by each ad-blocking blacklist is significantly similar (at least 65\% in 2017) to the annual set of a\&t domains from the preceding years, new a\&t domains are still appearing on the Alexa top 5K websites. %

\begin{figure}[ht!]
    \centering
    \begin{minipage}{0.48\columnwidth}
        \includegraphics[scale=0.29, keepaspectratio]{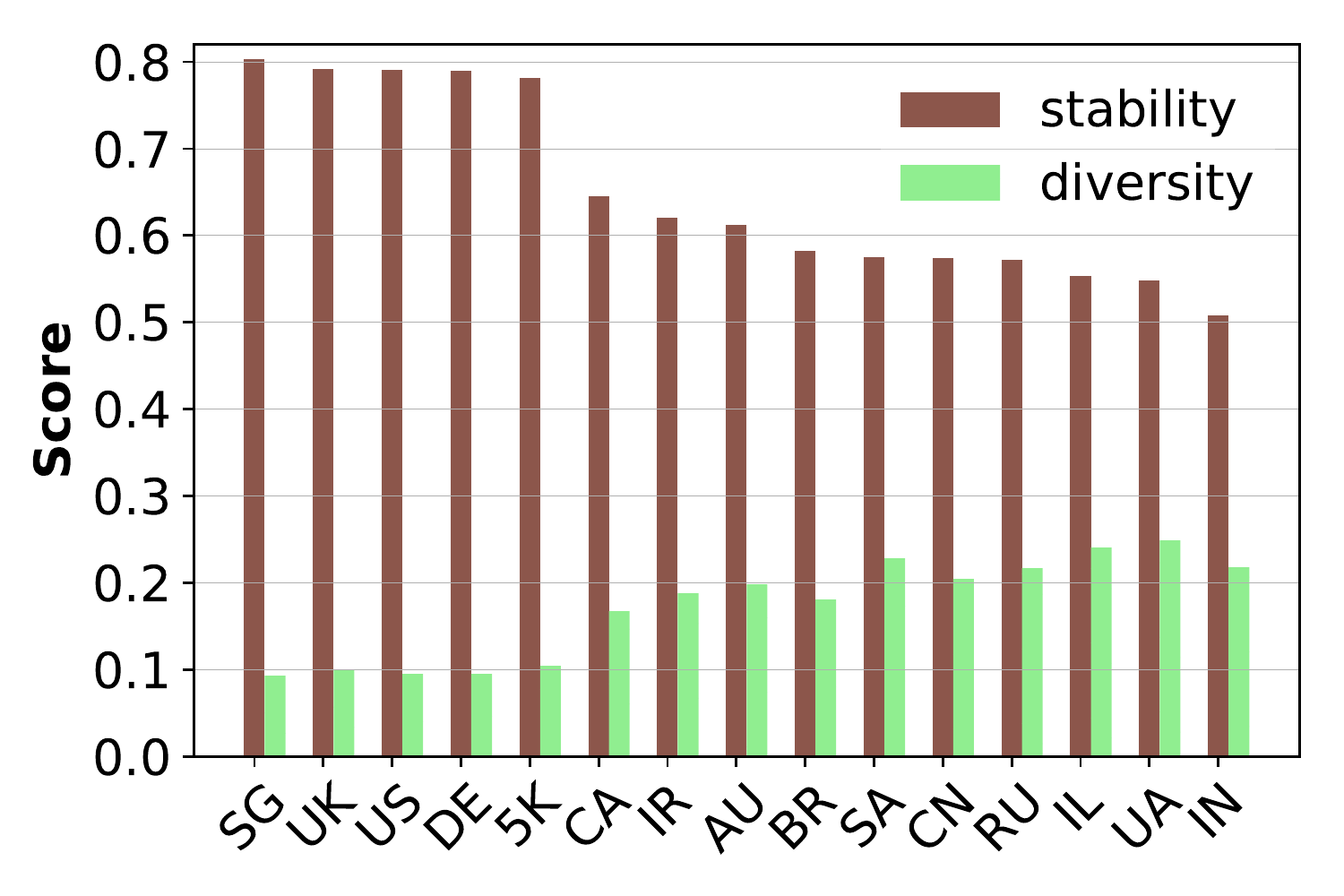}
        \subcaption{}
            \label{fig:5(a)}
    \end{minipage}
    \begin{minipage}{0.48\columnwidth}
        \includegraphics[scale=0.3, keepaspectratio]{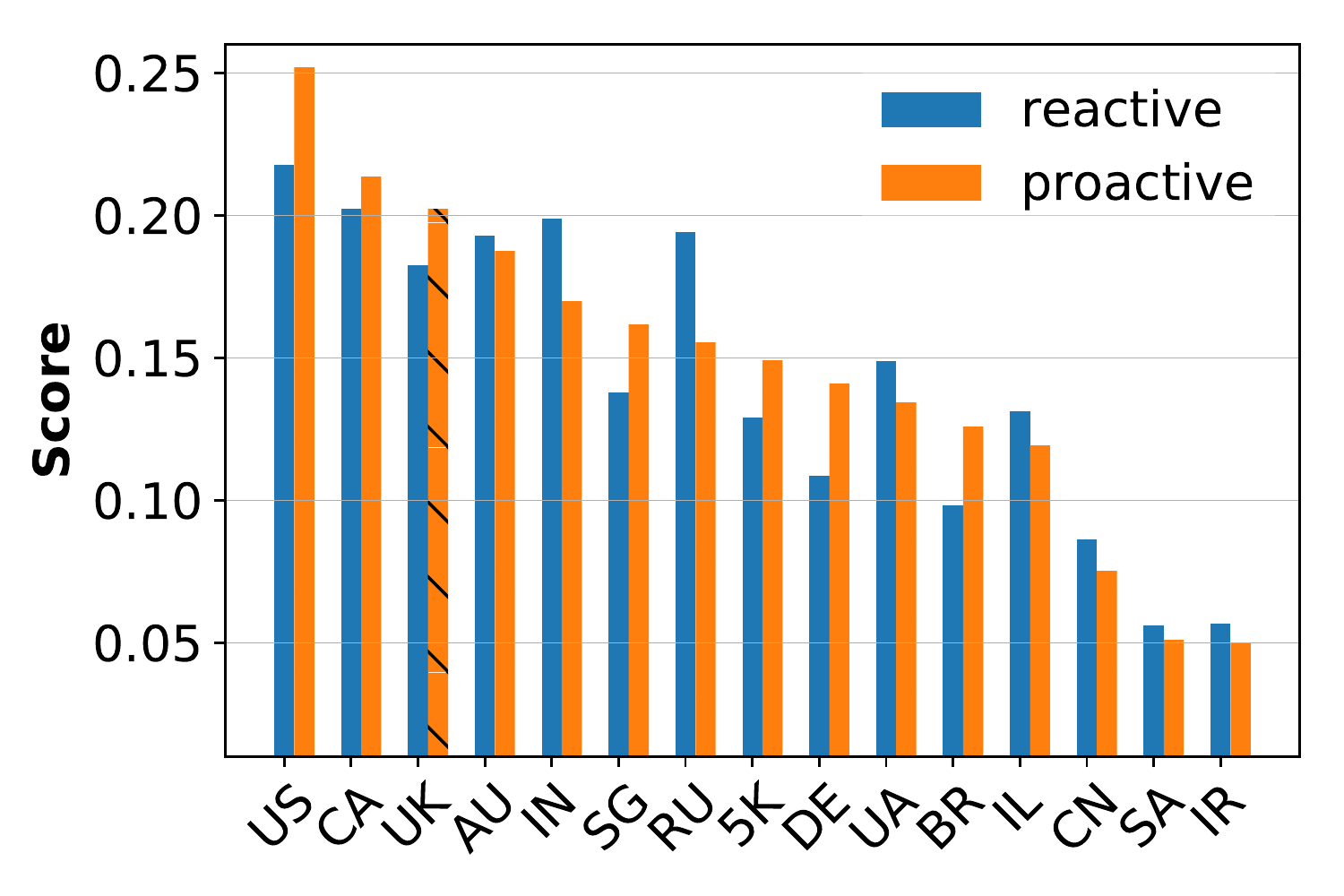}
        \subcaption{}
            \label{fig:5(b)}
    \end{minipage}
    \caption{\small (a) Average stability and diversity (b) Average reactive and proactive scores of countries. Countries with high proactive scores also tend to have a high number of a\&t domains on their Alexa top 5K websites (cf. Figure \ref{fig:1(b)}).}
    \label{fig:5}
\end{figure}

\textbf{Stability and diversity analysis of countries:} Figure \ref{fig:5(a)} shows the average stability and diversity scores of different countries based on the a\&t domains blocked on the Alexa top 5K websites. %
The score for each country is computed by measuring the stability and diversity for each year and then taking the average over the years. We observe that Singapore, UK, and US have the lowest diversity and conversely the highest stability scores, whereas Ukraine, Israel, and Saudi Arabia have the highest diversity and conversely the lowest stability scores. %

\begin{figure}[ht!]
\centering
\begin{minipage}{1.0\columnwidth}
\includegraphics[width=1.0\columnwidth]{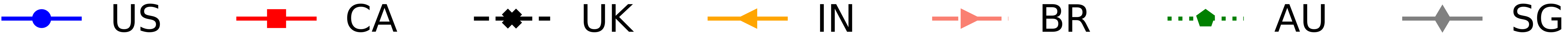}
\end{minipage}

\begin{minipage}{0.32\columnwidth}
\includegraphics[scale=0.2, keepaspectratio]{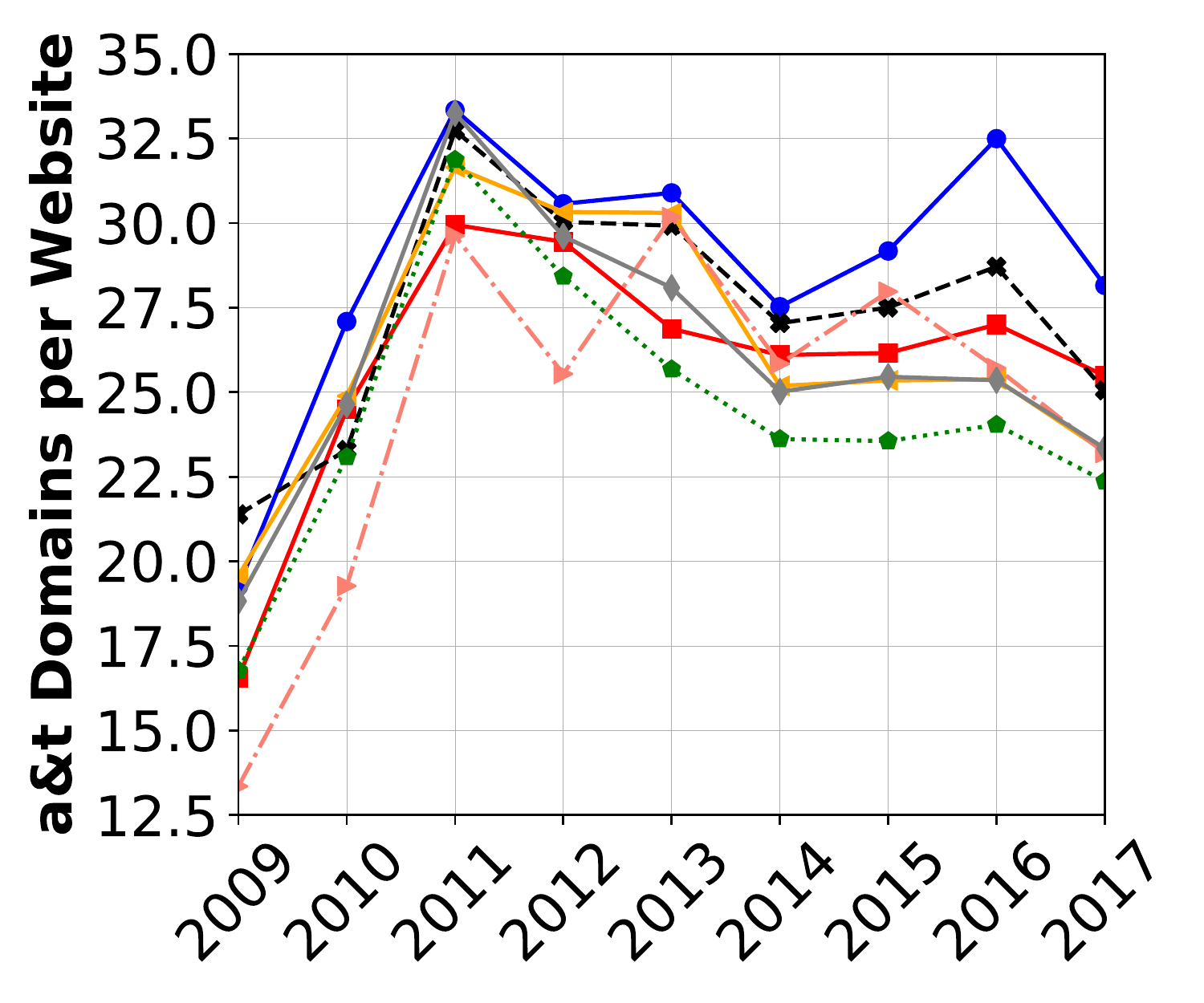}
\subcaption{}
    \label{fig:4(a)}
\end{minipage}
\begin{minipage}{0.32\columnwidth}
\includegraphics[scale=0.2, keepaspectratio]{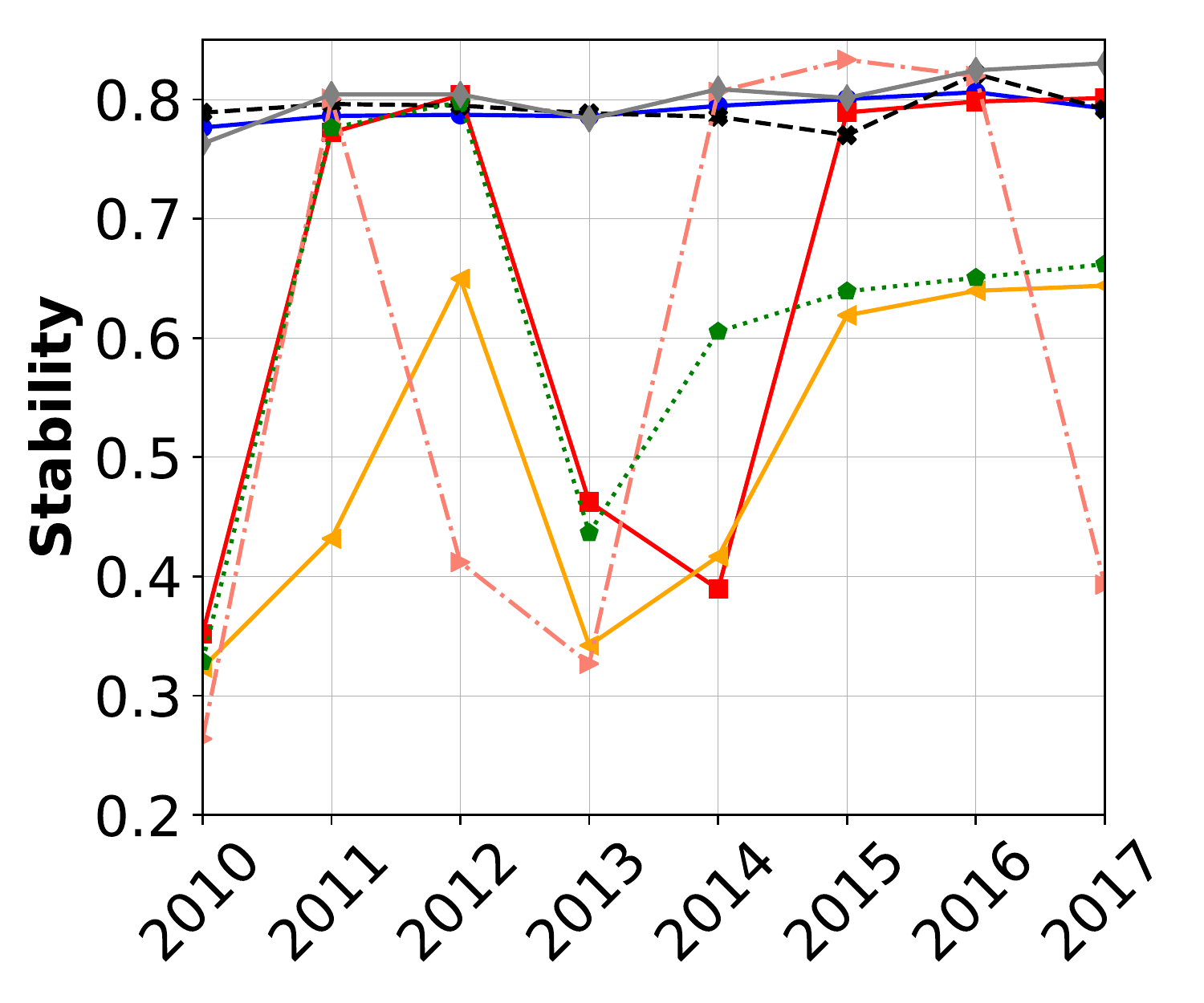}
\subcaption{}
    \label{fig:4(b)}
\end{minipage}
\begin{minipage}{0.32\columnwidth}
\includegraphics[scale=0.21, keepaspectratio]{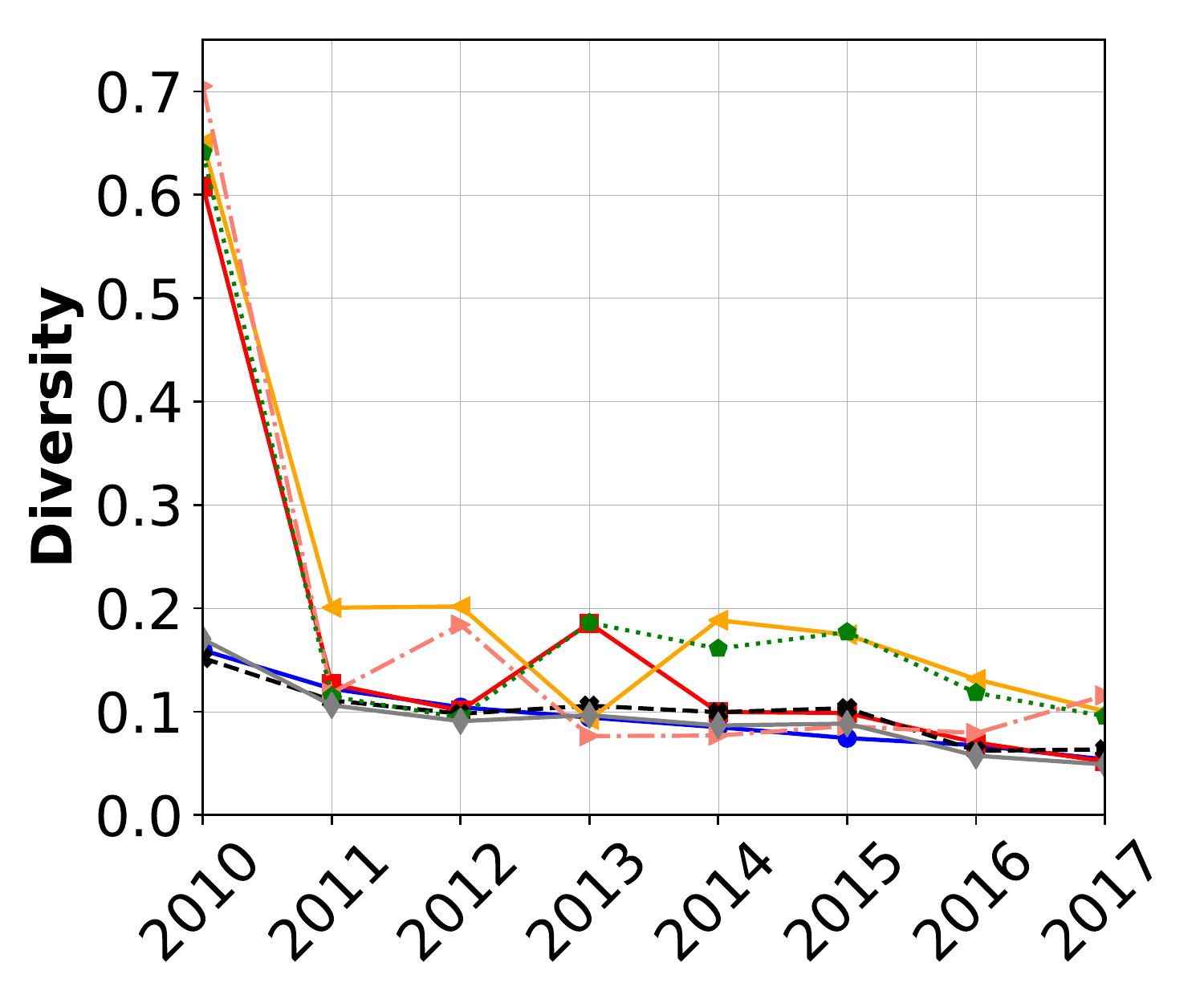}
\subcaption{}
    \label{fig:4(c)}
\end{minipage}
  \caption{\small (a) Annual a\&t domains per website blocked on the Alexa top 5K of countries (b) Average stability and (c) Average diversity scores of top seven countries.}
  \label{fig:4}
 
\end{figure}

Figures \ref{fig:4(b)} and \ref{fig:4(c)} show the annual stability and diversity scores, respectively, of the top seven countries (in Figure \ref{fig:1(b)}). Although the stability scores are fluctuating sharply for some countries, we notice that the diversity scores remain stable from 2011 onwards. This suggests that the percentage of new a\&t domains blocked on the Alexa top 5K websites in these countries drops slightly from 20\% in 2011 to 15\% in 2017 at the most.

\subsection{Update Speed}
\label{subsec:update}
\textbf{Reactive and proactive analysis of blacklists:} %
Figure \ref{fig:2(b)} shows the distribution of time taken in days to add a\&t domains observed on the Alexa top 5K websites, in the ad-blocking blacklists. In the distribution, we have omitted the a\&t domains that were observed on the Alexa top 5K websites before the ad-blocking blacklists' first snapshot. We notice that since we started capturing the a\&t domains in the ad-blocking blacklists (or the ad-blocking blacklists started their service), barring MVPs and CyberCrime, the ad-blocking blacklists have reported most of the a\&t domains proactively. 

\begin{figure}[!htb]
    \centering
    \begin{minipage}{0.3\columnwidth}
        \includegraphics[scale=0.32, keepaspectratio]{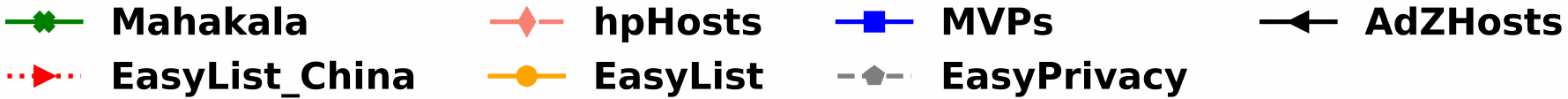}
    \end{minipage}
     
    \begin{minipage}{0.32\columnwidth}
        \includegraphics[scale=0.25, height=3.1cm, width=2.95cm]{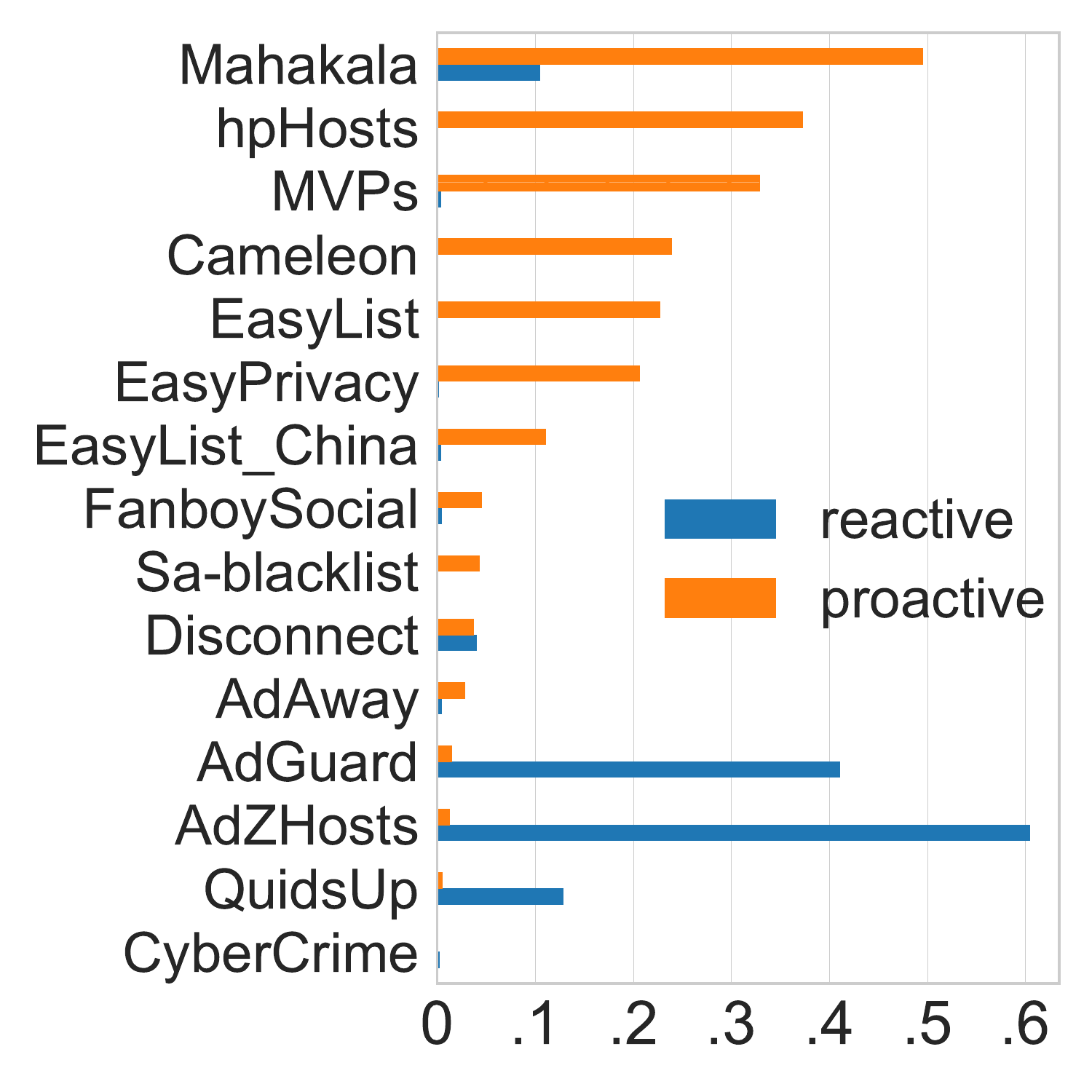}
        
        \subcaption{}
            \label{fig:7(a)}
    \end{minipage}
    \begin{minipage}{0.32\columnwidth}
        \includegraphics[scale=0.2, keepaspectratio]{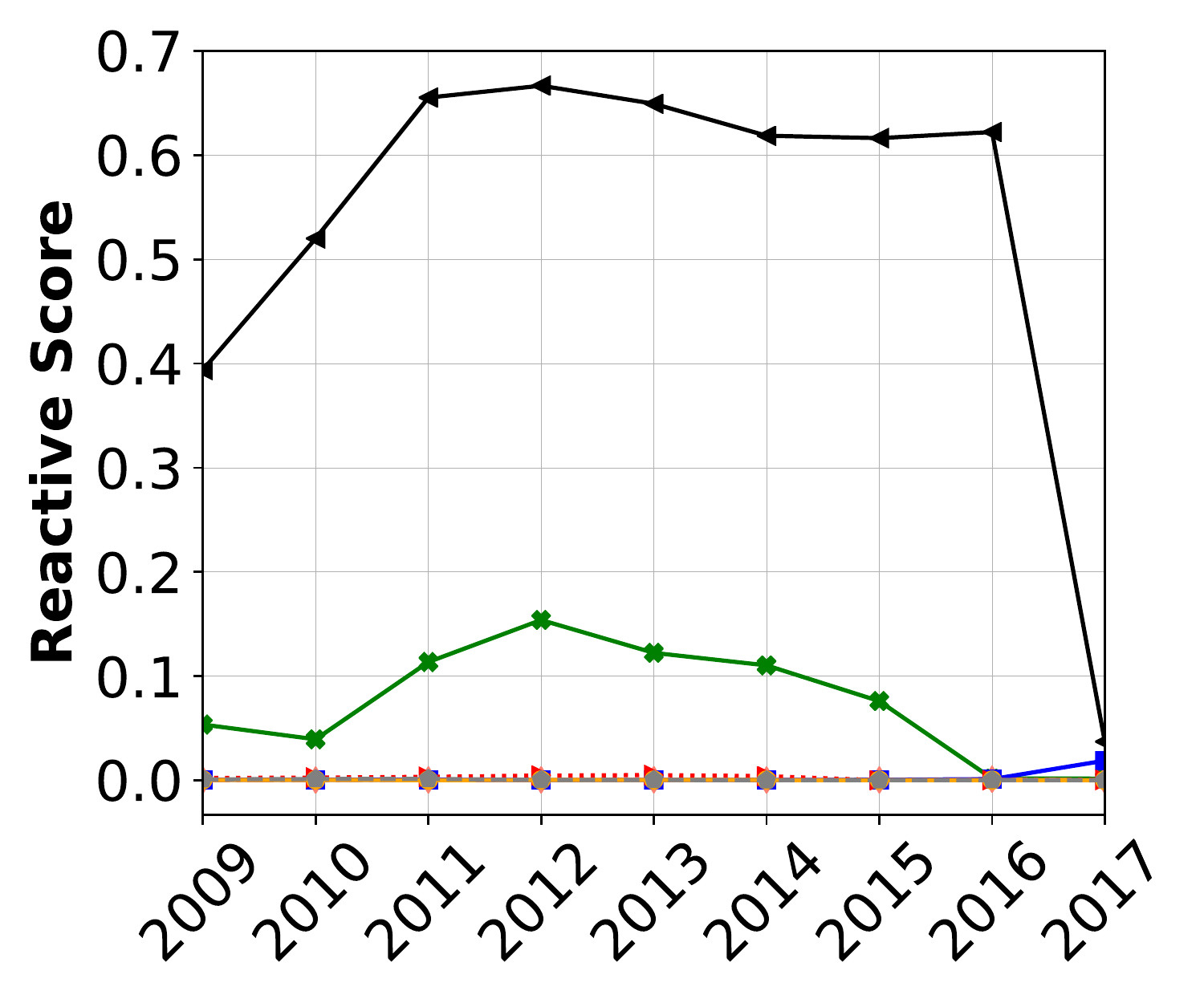}
        \subcaption{}
            \label{fig:7(b)}
    \end{minipage}
    \begin{minipage}{0.32\columnwidth}
        \includegraphics[scale=0.2, keepaspectratio]{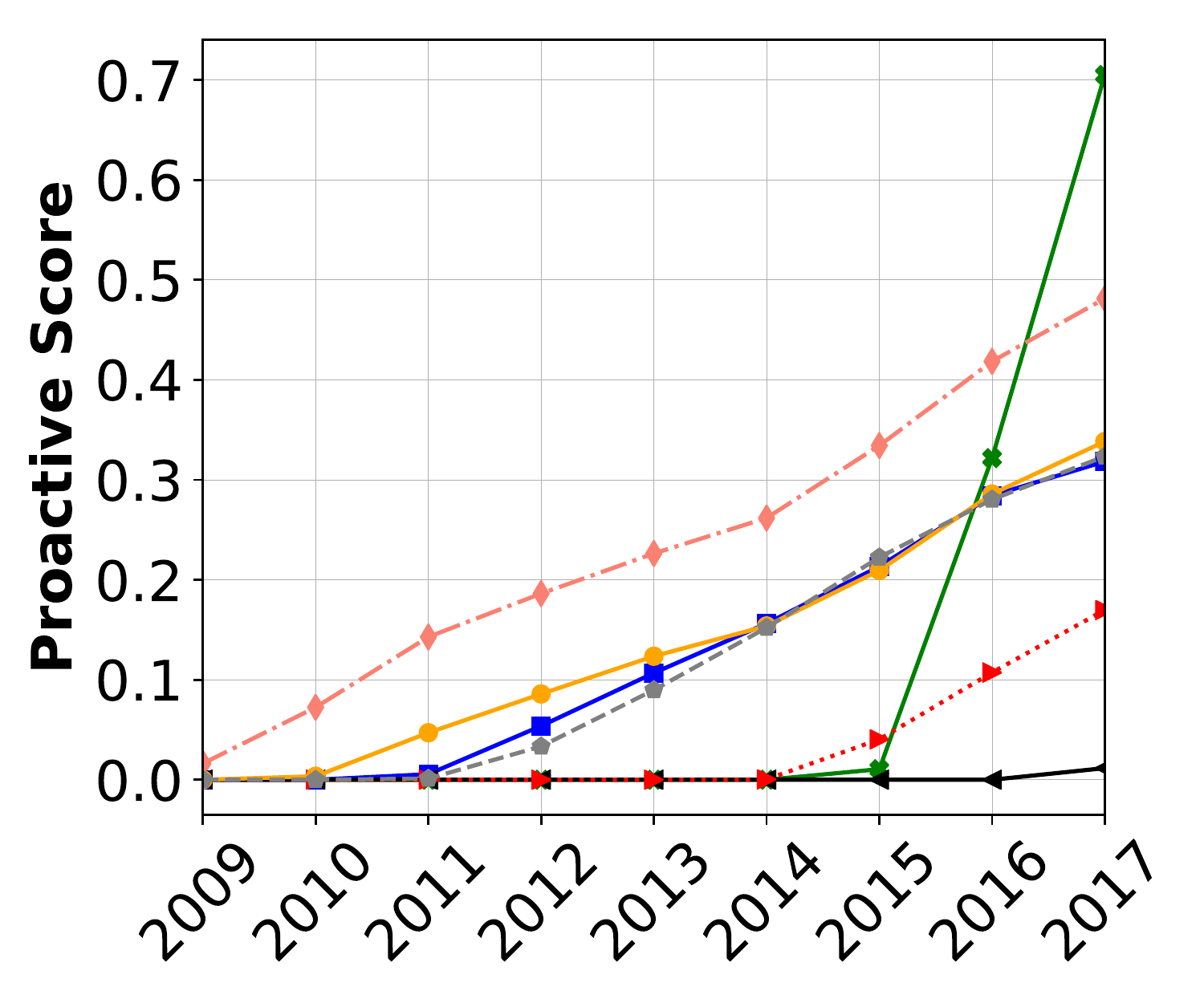}
        \subcaption{}
            \label{fig:7(c)}
    \end{minipage}
    \caption{\small (a) Average reactive and proactive scores of blacklists (b) Reactive and (c) Proactive scores of top seven blacklists.} 
    \label{fig:7}
\end{figure}

{\it Next}, we measure the update speed of ad-blocking blacklists with {\it reactive} and {\it proactive} metrics. Figure \ref{fig:7(a)} shows the average reactive and proactive scores for the ad-blocking blacklists. Mahakala has the highest average proactive score whereas AdZHosts has the highest average reactive score. We note that among the top seven blacklists in Table \ref{tab:blacklist_summary}, six are in the top seven blacklists by proactive score. 

Figure \ref{fig:7(b)} shows the annual reactive scores for the ad-blocking blacklists. All the a\&t domains blocked by AdZHosts from 2009 to 2016 were included in the first captured snapshot of AdZHosts. Therefore, the time difference was taken as $1$ day which resulted in a very high reactive score. The rise in the reactive score of AdZHosts from 2009 to 2011 is due to the rise in the number of a\&t domains blocked by AdZHosts in that period ({\it see} Figure \ref{fig:3(a)}). Mahakala's reactive score peaks in 2012, and then declines gradually from 2013 to 2015 because of an increase in the average time difference. Furthermore, in 2016, the reactive score of Mahakala decreases sharply because of the low number of reactive a\&t domains identified.

In Figure \ref{fig:7(c)} we notice a gradual increase in proactive score from the year in which we first captured the snapshot of the ad-blocking blacklist, because of an increase in the average time difference with each subsequent year. The proactive score does not rise significantly for AdZHosts in 2017 because the average time difference is least compared to the other ad-blocking blacklists. The sharp rise in Mahakala's proactive score in 2016 and 2017 is due to Mahakala blocking the most number of a\&t domains in the analyzed top 5K websites.
This suggests that since ad-blocking blacklists started their service, their proactive scores have been increasing gradually at nearly the same rate, apart from Mahakala which has risen sharply since 2015. Therefore, based on the number of a\&t domains blocked and the reactive and proactive score, we conclude that in the last two years, Mahakala has been the most effective ad-blocking blacklist. The number of black-listed domains in Mahakala are also the highest, and if we are applying ad-blocking blacklist on a performance constrained device then a leaner ad-blocking blacklist is most effective. In that case, EasyList, EasyList\_China and EasyPrivacy are the recommended ad-blocking blacklists.

\textbf{Reactive and proactive analysis of countries:} Figure \ref{fig:5(b)} illustrates the average reactive and proactive scores of different countries from 2009 to 2017. US has the highest reactive and proactive scores, whereas Iran has the lowest reactive and proactive scores. Out of the top seven countries ranked by the number of a\&t domains per website (in Figure \ref{fig:1(b)}), six have the highest proactive scores. Furthermore, among the bottom five countries ranked by a\&t domains per website, four are the bottom most countries by proactive scores. 

\begin{figure}[!htb]
    \centering
    \begin{minipage}{1.0\columnwidth}
    \includegraphics[width=1.0\columnwidth]{plots/cty_legend.pdf}
    \end{minipage}
    \begin{minipage}{0.48\columnwidth}
        \includegraphics[scale=0.295, keepaspectratio]{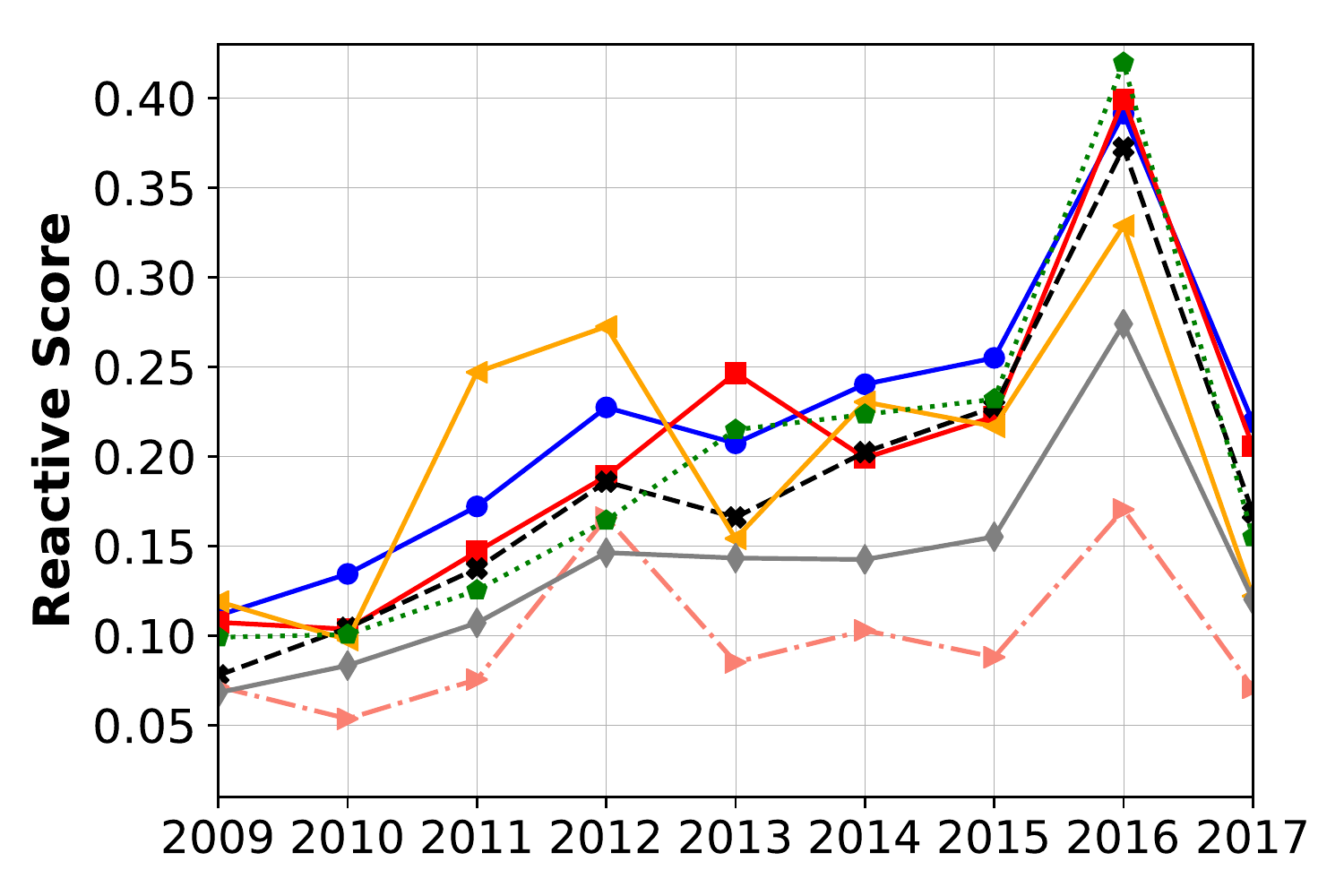}
        \subcaption{}
            \label{fig:6(a)}
    \end{minipage}
    \begin{minipage}{0.48\columnwidth}
        \includegraphics[scale=0.295, keepaspectratio]{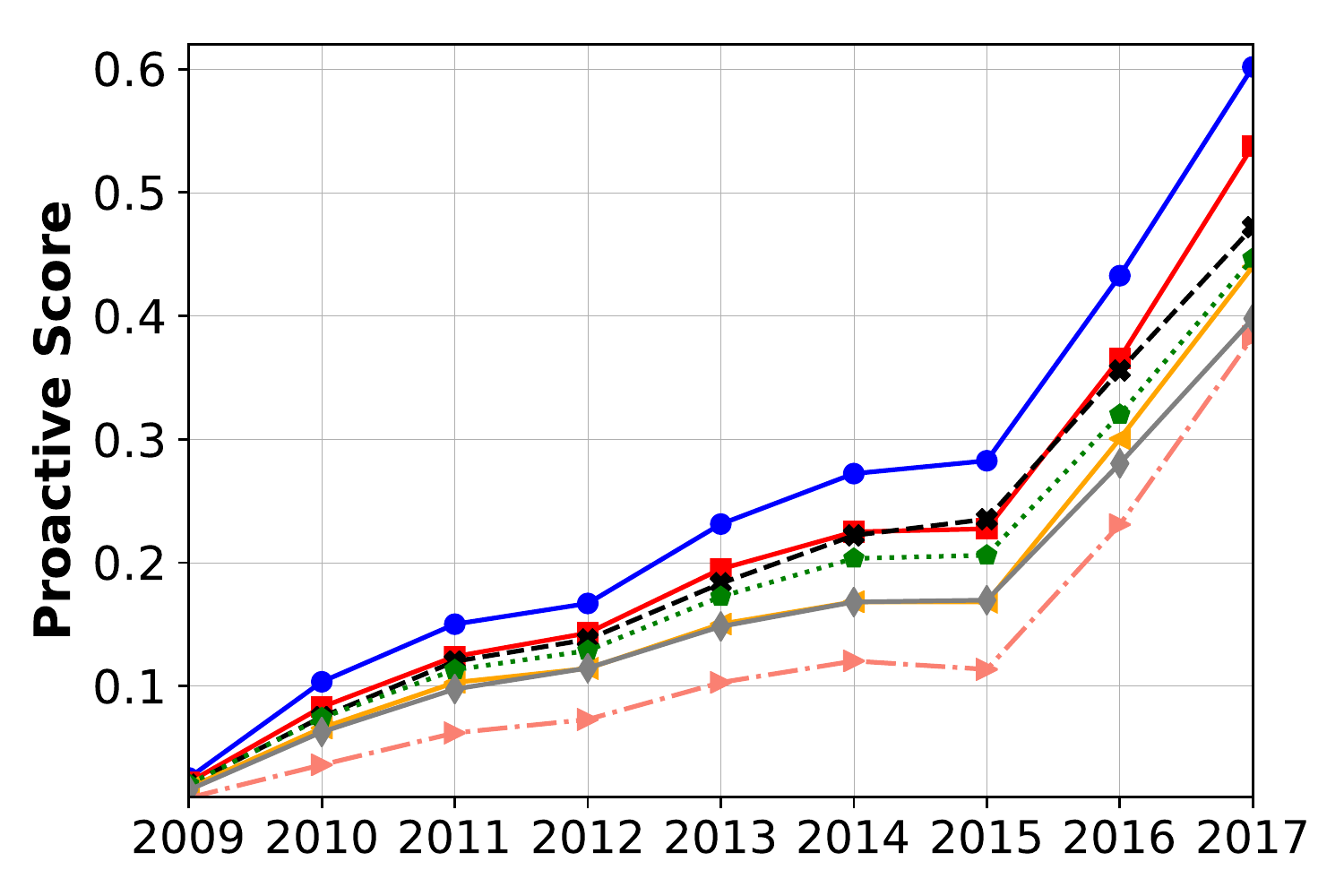}
        \subcaption{}
            \label{fig:6(b)}
    \end{minipage}
    \caption{\small (a) Reactive and (b) Proactive scores of countries. The number of a\&t domains detected reactively has decreased in recent years, whereas those identified proactively has risen.}%
    \label{fig:6}
\end{figure}

Figures \ref{fig:6(a)} and \ref{fig:6(b)} show the annual reactive and proactive scores, respectively, of the top seven countries by the number of a\&t domains per website. We notice a gradual increase in the reactive score of countries from 2009 to 2015. The reactive score depends on the number of distinct a\&t domains per website detected and the average time difference in days (cf Eq.~\ref{eq:react}). We notice a steep increase in reactive scores in 2016 because of less average time difference. Then in 2017, we observe a steep decline in reactive scores. Although the average time difference is less in 2017, the number of distinct a\&t domains per website detected is also less. In Figure \ref{fig:6(b)} we observe a gradual rise in proactive score from 2009 to 2015 due to an increase in average time difference (cf. Eq.~\ref{eq:proact}). Though the time difference increases steadily in 2016 and 2017, the number of a\&t domains blocked proactively in 2016 and 2017 is significantly greater than the previous years. Therefore, we observe a steep increase in proactive scores from 2016 onwards. This suggests that since 2016, the prevalence of a\&t domains detected proactively has significantly increased. 

\section{Related Work}
\label{sec:relatedWork}
Several research works focus on characterization and prevention of online tracking~\cite{Falahrastegar2014}, \cite{Ikram2017}, \cite{Lerner2016}, \cite{Krishnamurthy2009}, \cite{Mayer2012}, \cite{Metwalley2015}, \cite{Pujol2015}, \cite{Roesner2012a}, \cite{Roesner2012}. We classify the Web tracking literature in two classes and discuss them below.

\textbf{Prevalence of online ads and tracking:}
Krishnamurthy and Wills~\cite{Krishnamurthy2009}, 
use longitudinal measurement snapshots to show that the
third-party a\&t domains usage was already increasing significantly
between 2005 and 2008. Their work provides a comprehensive overview
of how some organizations (e.g., Google) have increased
their tracking coverage both by increased usage of some of
the third-party domains that they own, and through active
acquisition of new domains (e.g., DoubleClick) that provide
ads services, analytics, and tracking. %
Metwalley et al.,~\cite{Metwalley2015} did a passive study on the interaction of users with a\&t domains and ad-blockers. They showed that only one in five internet users have installed ad-blocker plugins on their web browsers, and nearly four out of five internet users interact with a\&t domains as soon as they begin to browse. Pujol et al.~\cite{Pujol2015} extended their work using traces by measuring traffic in ISP. They showed that the ad component is significant in web traffic, and many users install ad-blockers to block advertisements rather than protect their privacy. 

Our work focuses on the evolution of a\&t domains and ad-blocking lists. We perform a longitudinal study to show the prevalence of a\&t domains in the popular web of different countries. We show that websites embed new a\&t domains over time to evade ad-blocking lists and provide targeted advertisements to the users. The ad-blocking lists continue to detect new a\&t domains each year. %
We also provide the update speed of ad-blocking lists to illustrate if they have been reactive or proactive in handling a\&t domains. %

Lerner et al. found an increase in the prevalence and complexity of tracking, as well as an increase in the inter-connectedness of the ecosystem by analyzing Internet Archive data from 1996 to 2016~\cite{Lerner2016}. Fruchter et al. studied geographic variations in tracking~\cite{FruchterMSB15}. More recently, Libert studied third-party HTTP requests on the top 1 million sites~\cite{Libert15}, providing a view of tracking across the web. In this study, Libert showed that Google can track users across nearly 80\% of sites through its various third-party domains. Englehardt and Narayanan detected third-party fingerprinting services on Alexa top 1M websites~\cite{englehardt2016census}. 

\textbf{Characterization of online ads and tracking:} 
Roesner et al.,~\cite{Roesner2012} categorize mainstreaming tracking and analytics services and studied their prevalence in the wild. 
Ikram et al.,~\cite{Ikram2017} evaluate the usability of five different tracking prevention plugins and reveal that contemporary blacklists based tracking prevention tools are inefficient and have high false positives. Ikram et al.~\cite{ikram2017first} classify the tools for browser plugins, VPN or proxy-based tools, DNS-based filtering, and stand-alone browsers. Their work also delineate tracking prevention techniques for mobile as well as desktop environments. Blacklists are discussed and machine-learning based techniques are evaluated for tracking prevention~\cite{Ikram2017, Bhagavatula2014}.

\section{Limitations and Concluding Remarks}
\label{sec:conc}
 Despite our best efforts to collect the dataset from the Wayback Machine to perform our study, there are still some limitations. 
 First, the webpage requested to Wayback Machine sometimes encounters a 301 redirect, meaning that the page has permanently moved to a new location, or a 302 redirect meaning that the move is only temporary.  
Second, websites are archived with varied frequency on Wayback Machine.  
Hence, it is possible to miss an a\&t domain embedded in a website for the duration the website was not archived by Wayback Machine. 
When we encounter a {\tt web page redirect} on a given snapshot, we discard that snapshot and pick another snapshot from the same three month interval. Similarly, if we are unable to obtain any snapshot in that three month interval after four tries, we skip that interval altogether and move to the next three month interval. For a website, if we are unable to capture at least a single snapshot for any year between 2009 and 2017, we discard that website from our analysis.

We have presented an approach to measure the longitudinal performance of ad-blocking blacklists. We have proposed two metrics, stability and diversity, that measure the rate of change of ad-blocking blacklists, and introduced two metrics, reactive and proactive, that measure the update speed of ad-blocking blacklists. Our analysis revealed that since 2011, more than 50\% of the crawled websites embedded a different number of a\&t domains compared to the preceding year. Ad-blocking blacklists also continue to detect new a\&t domains each year (at most 15\% in 2017). Among the ad-blocking blacklists, Mahakala and hpHosts blocked the most number of a\&t domains on the web, but these two ad-blocking blacklists also contain the highest number of black-listed domains. On resource-constrained devices, leaner ad-blocking blacklists like EasyList, EasyPrivacy, and EasyList\_China, that provide slightly less coverage are more suitable.  
Among the countries we evaluated, the US, Canada, and the UK have the highest number of distinct a\&t domains per website in the popular web and also the highest proactive scores. This suggests that the ad-blocking lists update by prioritizing the a\&t domains reported in the popular websites from these countries.

\balance
\small

\bibliographystyle{IEEEtranS}

\bibliography{adblock_main}

\begin{thebibliography}{10}
\providecommand{\url}[1]{#1}
\csname url@samestyle\endcsname
\providecommand{\newblock}{\relax}
\providecommand{\bibinfo}[2]{#2}
\providecommand{\BIBentrySTDinterwordspacing}{\spaceskip=0pt\relax}
\providecommand{\BIBentryALTinterwordstretchfactor}{4}
\providecommand{\BIBentryALTinterwordspacing}{\spaceskip=\fontdimen2\font plus
\BIBentryALTinterwordstretchfactor\fontdimen3\font minus
  \fontdimen4\font\relax}
\providecommand{\BIBforeignlanguage}[2]{{%
\expandafter\ifx\csname l@#1\endcsname\relax
\typeout{** WARNING: IEEEtranS.bst: No hyphenation pattern has been}%
\typeout{** loaded for the language `#1'. Using the pattern for}%
\typeout{** the default language instead.}%
\else
\language=\csname l@#1\endcsname
\fi
#2}}
\providecommand{\BIBdecl}{\relax}
\BIBdecl

\bibitem{adblock}
``{Adblock: content filtering and ad blocking browser extension},''
  \url{https://getadblock.com}.

\bibitem{adblockplus}
``{Adblock Plus: open-source browser extension for content-filtering and ad
  blocking},'' \url{https://adblockplus.org}.

\bibitem{CC_stats}
``{CyberCrime Statistics},'' \url{https://cybercrime-tracker.net/stats.php}.

\bibitem{ghostery}
``{Ghostery: open-source browser extension and mobile browser application to
  control JavaScript tags},'' \url{https://www.ghostery.com}.

\bibitem{wayback}
``{Internet Archive: Digital Library of Free \& Borrowable Books, Movies, Music
  \& Wayback Machine},'' \url{http://archive.org}.

\bibitem{privacybadger}
``{Privacy Badger: open-source browser extension for blocking advertisements
  and tracking cookies},'' \url{https://www.eff.org/fr/node/99095}.

\bibitem{disconnect}
``{Canonical repository for the Disconnect services file},''
  \url{https://disconnect.me/trackerprotection}, 2018.

\bibitem{adsnetworksmalware}
``{Dasient Smart Web Security Q3 2010 Malware Update},''
  \url{https://web-beta.archive.org/web/20111017194759/http://www.dasient.com/documents/Dasient_3Q_2010_Qtrly_Malware_F.pdf},
  2018.

\bibitem{adaway}
``{AdAway Hosts},'' \url{https://adaway.org/hosts.txt}, 2019.

\bibitem{adguard}
``Ad{G}uard,'' \url{https://adguard.com}, 2019.

\bibitem{adzhosts}
``{Ad{ZH}osts},'' \url{https://adzhosts.eu}, 2019.

\bibitem{cameleon}
``Cameleon,'' \url{http://sysctl.org/cameleon}, 2019.

\bibitem{cybercrime}
``Cybercrime-tracker,'' \url{https://cybercrime-tracker.net}, 2019.

\bibitem{easylist}
``{EasyList},'' \url{https://easylist.to/easylist/easylist.txt}, 2019.

\bibitem{EasyListChina}
``{EasyList China+EasyList},''
  \url{https://easylist-downloads.adblockplus.org/easylistchina+easylist.txt},
  2019.

\bibitem{easyprivacy}
``{EasyPrivacy},'' \url{https://easylist.to/easylist/easyprivacy.txt}, 2019.

\bibitem{fanoby}
``{Fanboy's Social Blocking List},''
  \url{https://easylist-downloads.adblockplus.org/fanboy-social.txt}, 2019.

\bibitem{hpHost}
``{hpHosts Online -- Simple, Searchable \& FREE!}''
  \url{http://www.hosts-file.net}, 2019.

\bibitem{mahakala}
``{Mahakala Adblocking Hosts},'' \url{https://adblock.mahakala.is/}, 2019.

\bibitem{momento}
``{Momento: Time Travel},'' \url{http://timetravel.mementoweb.org}, 2019.

\bibitem{mvpshosts}
``{MVPs Hosts Lists},'' \url{http://winhelp2002.mvps.org/hosts.txt}, 2019.

\bibitem{pihole}
``{Pi-hole: A black hole for Internet advertisements},''
  \url{https://pi-hole.net}, 2019.

\bibitem{privoxy}
``{Privoxy - Home Page},'' \url{https://www.privoxy.org}, 2019.

\bibitem{proxmitron}
``{Proxomitron.Info ...the Webhiker's Guide to Proxomitron.}''
  \url{https://proxomitron.info}, 2019.

\bibitem{QuidsUp}
``Quids{U}p,'' \url{https://quidsup.net}, 2019.

\bibitem{sa-filterlist}
``Sa-blacklist,'' \url{http://www.stearns.org/sa-blacklist}, 2019.

\bibitem{barth2009securing}
A.~Barth, C.~Jackson, and J.~C. Mitchell, ``Securing frame communication in
  browsers,'' \emph{Communications of the ACM}, 2009.

\bibitem{Bhagavatula2014}
S.~Bhagavatula, C.~Dunn, C.~Kanich, M.~Gupta, and B.~Ziebart, ``Leveraging
  machine learning to improve unwanted resource filtering,'' in
  \emph{Proceedings of the 2014 Workshop on Artificial Intelligent and Security
  Workshop}.\hskip 1em plus 0.5em minus 0.4em\relax ACM, 2014, pp. 95--102.

\bibitem{englehardt2016census}
S.~Englehardt and A.~Narayanan, ``Online tracking: A 1-million-site measurement
  and analysis,'' in \emph{Proceedings of ACM CCS 2016}, 2016.

\bibitem{Falahrastegar2014}
M.~Falahrastegar, H.~Haddadi, S.~Uhlig, and R.~Mortier, ``The rise of
  panopticons: Examining region-specific third-party web tracking.'' in
  \emph{TMA}, 2014, pp. 104--114.

\bibitem{FruchterMSB15}
\BIBentryALTinterwordspacing
N.~Fruchter, H.~Miao, S.~Stevenson, and R.~Balebako, ``Variations in tracking
  in relation to geographic location,'' \emph{CoRR}, vol. abs/1506.04103, 2015.
  [Online]. Available: \url{http://arxiv.org/abs/1506.04103}
\BIBentrySTDinterwordspacing

\bibitem{Ikram2017}
M.~Ikram, H.~J. Asghar, M.~A. Kaafar, A.~Mahanti, and B.~Krishnamurthy,
  ``Towards seamless tracking-free web: Improved detection of trackers via
  one-class learning,'' \emph{Proceedings on Privacy Enhancing Technologies},
  vol. 2017, no.~1, pp. 79--99, 2017.

\bibitem{ikram2017first}
M.~Ikram and M.~A. Kaafar, ``A first look at mobile ad-blocking apps,'' in
  \emph{Network Computing and Applications (NCA), 2017 IEEE 16th International
  Symposium on}.\hskip 1em plus 0.5em minus 0.4em\relax IEEE, 2017, pp. 1--8.

\bibitem{ikram2019chain}
M.~Ikram, R.~Masood, G.~Tyson, M.~A. Kaafar, N.~Loizon, and R.~Ensafi, ``The
  chain of implicit trust: An analysis of the web third-party resources
  loading,'' in \emph{WWW}, 2019.

\bibitem{Krishnamurthy2009}
B.~Krishnamurthy and C.~Wills, ``Privacy diffusion on the web: a longitudinal
  perspective,'' in \emph{Proceedings of the 18th international conference on
  World wide web}.\hskip 1em plus 0.5em minus 0.4em\relax ACM, 2009, pp.
  541--550.

\bibitem{Lerner2016}
A.~Lerner, A.~K. Simpson, T.~Kohno, and F.~Roesner, ``Internet jones and the
  raiders of the lost trackers: An archaeological study of web tracking from
  1996 to 2016,'' in \emph{USENIX Security}, 2016.

\bibitem{Libert15}
\BIBentryALTinterwordspacing
T.~Libert, ``Exposing the hidden web: An analysis of third-party {HTTP}
  requests on 1 million websites,'' \emph{CoRR}, vol. abs/1511.00619, 2015.
  [Online]. Available: \url{http://arxiv.org/abs/1511.00619}
\BIBentrySTDinterwordspacing

\bibitem{Mayer2012}
J.~R. Mayer and J.~C. Mitchell, ``Third-party web tracking: Policy and
  technology,'' in \emph{Security and Privacy (SP), 2012 IEEE Symposium
  on}.\hskip 1em plus 0.5em minus 0.4em\relax IEEE, 2012, pp. 413--427.

\bibitem{Metwalley2015}
H.~Metwalley, S.~Traverso, M.~Mellia, S.~Miskovic, and M.~Baldi, ``The online
  tracking horde: a view from passive measurements,'' in \emph{International
  Workshop on Traffic Monitoring and Analysis}.\hskip 1em plus 0.5em minus
  0.4em\relax Springer, 2015, pp. 111--125.

\bibitem{Pujol2015}
E.~Pujol, O.~Hohlfeld, and A.~Feldmann, ``Annoyed users: Ads and ad-block usage
  in the wild,'' in \emph{Proceedings of the 2015 ACM Conference on Internet
  Measurement Conference}.\hskip 1em plus 0.5em minus 0.4em\relax ACM, 2015,
  pp. 93--106.

\bibitem{Roesner2012}
F.~Roesner, T.~Kohno, and D.~Wetherall, ``Detecting and defending against
  third-party tracking on the web,'' in \emph{Proceedings of the 9th USENIX
  conference on Networked Systems Design and Implementation}.\hskip 1em plus
  0.5em minus 0.4em\relax USENIX Association, 2012, pp. 12--12.

\bibitem{Roesner2012a}
F.~Roesner, C.~Rovillos, T.~Kohno, and D.~Wetherall, ``Sharemenot: Balancing
  privacy and functionality of third-party social widgets,'' \emph{Usenix;
  login}, 2012.

\bibitem{vastel2018filters}
A.~Vastel, P.~Snyder, and B.~Livshits, ``Who filters the filters: Understanding
  the growth, usefulness and efficiency of crowdsourced ad blocking,''
  \emph{arXiv preprint arXiv:1810.09160}, 2018.

\bibitem{wills2016ad}
C.~E. Wills and D.~C. Uzunoglu, ``What ad blockers are (and are not) doing,''
  in \emph{HotWeb}, 2016.

\bibitem{zhu2018measuring}
S.~Zhu, X.~Hu, Z.~Qian, Z.~Shafiq, and H.~Yin, ``Measuring and disrupting
  anti-adblockers using differential execution analysis,'' in \emph{NDSS},
  2018.

\end{thebibliography}

\end{document}